\documentclass[fleqn,usenatbib,useAMS]{mnras}

\usepackage{graphicx}	
\usepackage{amsmath}	
\usepackage{amssymb}	
\usepackage{multicol}        
\usepackage{bm}		
\usepackage{pdflscape}	
\usepackage{makecell}

\newcommand{\propane}{{\sc ProPane}}
\newcommand{\protools}{{\sc ProTools}}
\newcommand{\profuse}{{\sc ProFuse}}	
\newcommand{\profound}{{\sc ProFound}}	
\newcommand{\profit}{{\sc ProFit}}
\newcommand{\prospect}{{\sc ProSpect}}
\newcommand{\Rfits}{{\sc Rfits}}
\newcommand{\Rwcs}{{\sc Rwcs}}

\newcommand{\CFITSIO}{{\sc CFITSIO}}
\newcommand{\wcslib}{{\sc wcslib}}

\newcommand{\R}{{\sc R}}
\newcommand{\C}{{\sc C}}
\newcommand{\Cpp}{{\sc C++}}
\newcommand{\CImg}{{\sc CImg}}
\newcommand{\FITS}{{\sc FITS}}
\newcommand{\HDF}{{\sc HDF5}}
\newcommand{\imager}{{\sc imager}}
\newcommand{\swarp}{{\sc SWarp}}
\newcommand{\drizzle}{{\sc Drizzle}}

\usepackage[T1]{fontenc}
\usepackage{ae,aecompl}

\usepackage{newtxtext,newtxmath}

\title[ProPane: Image Warping with Fire]{ProPane: Image Warping with Fire}

\author[A.~S.~G.~ Robotham et al.]{
A.~S.~G. Robotham,$^{1}$\thanks{E-mail: aaron.robotham@uwa.edu.au}
R. Tobar,$^{1}$
S. Bellstedt,$^{1}$
S. Casura,$^{2}$
R.~H.~W. Cook,$^{1}$
J.~C.~J. D'Silva,$^{1}$
\newauthor
L.~J. Davies,$^{1}$
S.~P. Driver,$^{1}$
J. Li,$^{1}$
L.~P. Garate-Nuñez$^{1}$
\\\\
$^{1}$ICRAR, M468, University of Western Australia, Crawley, WA 6009, Australia \\
$^{2}$Hamburger Sternwarte, UniversitÃ€t Hamburg, Gojenbergsweg 112, D-21029 Hamburg, Germany
}

\date{Last updated YYYY MM DD; in original form YYYY MM DD}

\pubyear{2023}

\begin{document}
\label{firstpage}
\pagerange{\pageref{firstpage}--\pageref{lastpage}}
\maketitle

\begin{abstract}
In this paper we introduce the software package \propane, written for the \R{} data analysis language. \propane{} combines the full range of \wcslib{} projections with the \Cpp{} image manipulation routines provided by the \CImg{} library. \propane{} offers routines for image warping and combining (including stacking), and various related tasks such as image alignment tweaking and pixel masking. It can stack an effectively unlimited number of target frames using multiple parallel cores, and offers threading for many lower level routines. It has been used for a number of current and upcoming large surveys, and we present a range of its capabilities and features. \propane{} is already available under a permissive open-source LGPL-3 license at github.com/asgr/ProPane (DOI: 10.5281/zenodo.10057053).
\end{abstract}




\section{Introduction}

One of the foundational tasks of modern astronomy surveys is combining multiple observations of imaging data over an area of sky \citep[e.g.\ see][]{2002PASP..114..144F, 2007ApJS..172..196K, 2014PASP..126..158G}. When this is done largely to extend the area it is often referred to as `mosaicking', and if this is done to increase the apparent depth of a region of sky it is called `stacking'. Algorithmically, they are much the same process, and in the software presented here we largely think of both operations as a form of image `combining'. Combining images involves potentially a large number of preparatory steps, e.g.\ data alignment; background subtraction; and artefact masking are usually the minimum steps required, but occasionally additional homogenisation of the point spread function is also desired. \propane{} is a new software package that aims to automate most of these tasks, along with the final combining process we wish to carry out.

\propane{} is a component of the \protools{} software suite \citep{ProTools}\footnote{github.com/asgr/ProTools}, and is usually the step run before extracting photometry with \profound{} \citep{2018MNRAS.476.3137R}, and fitting spectral energy distributions (SEDs) with \prospect{} \citep{2020MNRAS.495..905R}. The resulting images are usually also the inputs we desire for potential structural decomposition with \profit{} \citep{2017MNRAS.466.1513R} and \profuse{} \citep{2022MNRAS.513.2985R}. In this paper we briefly discuss the lower level \protools{} software relevant to \propane{} (Section \ref{sec:software}); discuss the algorithmic choices made when writing \propane{} itself (Section \ref{sec:propane}); discuss its recent applications in the literature (Section \ref{sec:applications}); and finally conclude and reflect on future use cases and development directions (Section \ref{sec:conclusions}). Because of the large number of software packages and astronomy acronyms used in this paper, a compact glossary is included in the Appendix.

\section{Foundational Software}
\label{sec:software}

\propane{} is written in the \R{} statistical computing language \citep{citeR}. By necessity \propane{} builds on lower-level software packages that provide some basic utility. Below we briefly summarise some key external \R{} packages, and what functionality they bring to \propane.

\subsection{Rfits}

\Rfits{} is a mid- to high-level interface to the \CFITSIO{} \Cpp{} software library that allows efficient interaction with the hierarchical \FITS{} image and table storage format \citep{1999ASPC..172..487P}\footnote{heasarc.gsfc.nasa.gov/fitsio/}. \FITS{} is close to a standard format within the astronomical software ecosystem, but is broadly comparable in terms of functionality to some aspects of the more broadly popular \HDF{} format\footnote{Hierarchical Data Format 5; www.hdfgroup.org/solutions/hdf5/}. \Rfits{} has existed as a public GitHub repository\footnote{github.com/asgr/Rfits} since September 2019, and has been continuously developed and improved since then. It is quite widely used in its current v1.10 form, and considered largely stable in terms of features and performance. It is comprehensively documented with a 48 page feature-complete manual, passes all current Central \R{} Archive Network (CRAN) package consistency checks, and has a large number of online examples and vignettes covering the majority of its capabilities\footnote{rpubs.com/asgr}. As such, we only briefly mention the key features relevant for \propane{} here. \CFITSIO{} v3.58 is included as part of the \Rfits{} package, so no alternative downloading or linking to this library is necessary.

\Rfits{} offers a range of basic read and write capability for images (including 1--4D vectors / images / cubes / arrays) and tables (binary and ascii), including the ability to read all extensions of a given \FITS{} file into one large \R{} S3 object list structure (one of the most general purpose in-memory data formats available in \R). It also offers pointer methods to \FITS{} images, and a large number of methods to allow for many inspection operations without fully loading the pixel data. A non-exhaustive list includes being able to interrogate the image dimensions; central coordinates; corner coordinates; central pixel scale; and central pixel area. \Rfits{} also supports all paradigms of image compression available in \CFITSIO, both for floating point and integer data.

Key to its utility within \propane, \Rfits{} also offers efficient methods to use pointer objects to access only pixel subsets of images. This means without loading the entire (often large) image into memory, the methods available can be combined to identify which image pixels overlap with the target world coordinate system (WCS) we wish to warp onto. Only these pixels are then loaded into memory, with their relevant WCS information being dynamically modified to ensure pixels are still correctly positioned in terms of their celestial coordinates.

\Rfits{} fully supports the whole range of \FITS{} key words, history and comments. Whilst the \FITS{} standard specifies how the data must be represented on-disk, there are clearly a lot of mechanisms to represent a \FITS{} file in memory, depending on the flexibility of the language interfacing with it. In the case of \R, there is a lot of flexibility in how we organise a hierarchical format such as \FITS, however there are restrictions too. \R{} notably lacks a native single precision floating point format, in fact it only supports double precision floating points number and 32 bit integers. Because the characteristic (integer) component of double precision numbers can be used to represent integers, \R{} effectively supports integers of magnitude $2^{52}$. The {\sc bit64} package gives users access to full 64 bit integers within both \R{} and \Rfits, if necessary.

These restrictions mean the vast range of number formats available in \CFITSIO{} have to be mapped onto their most appropriate \R{} representation, and when writing out to file users have to be aware of the written format type (if not directly associated with a native \R{} type), and specify the desired format type if necessary. In practice, the default behaviour captures the typical use case of loading \FITS{} data, interacting with it in \R, and writing it back out to disk. However, it will often be the case that files will not be identical at the bit level due to the internal conversions necessary (i.e.\ there may be some machine precision error).

\subsection{Rwcs}

Whilst \Rfits{} contains methods to dynamically interact and update the WCS of a target \FITS{} object, it does not provide utilities to convert pixel positions to celestial coordinate positions. For this we need to interface with a WCS library, where the \Rwcs{} package is an interface to the popular and well maintained \wcslib{} \C{} library \citep{2002AandA...395.1077C}\footnote{www.atnf.csiro.au/people/mcalabre/WCS/}. \Rwcs{} has existed as a public GitHub repository since November 2019\footnote{github.com/asgr/Rwcs}, being developed in close union with the \Rfits{} package. It is used for a number of plotting and projection tasks by the astronomy community in its current v1.7 form, and considered largely stable in terms of features and performance. It is comprehensively documented with a 25 page feature-complete manual, passes all current CRAN package consistency checks, and has a large number of online examples and vignettes covering the majority of its capabilities\footnote{rpubs.com/asgr}. As such, we only briefly mention the key features relevant for \propane{} here. \wcslib{} v7.9 is included as part of the \Rwcs{} package, so no alternative downloading or linking to this library is necessary.

\begin{figure}
 \includegraphics[width=\columnwidth]{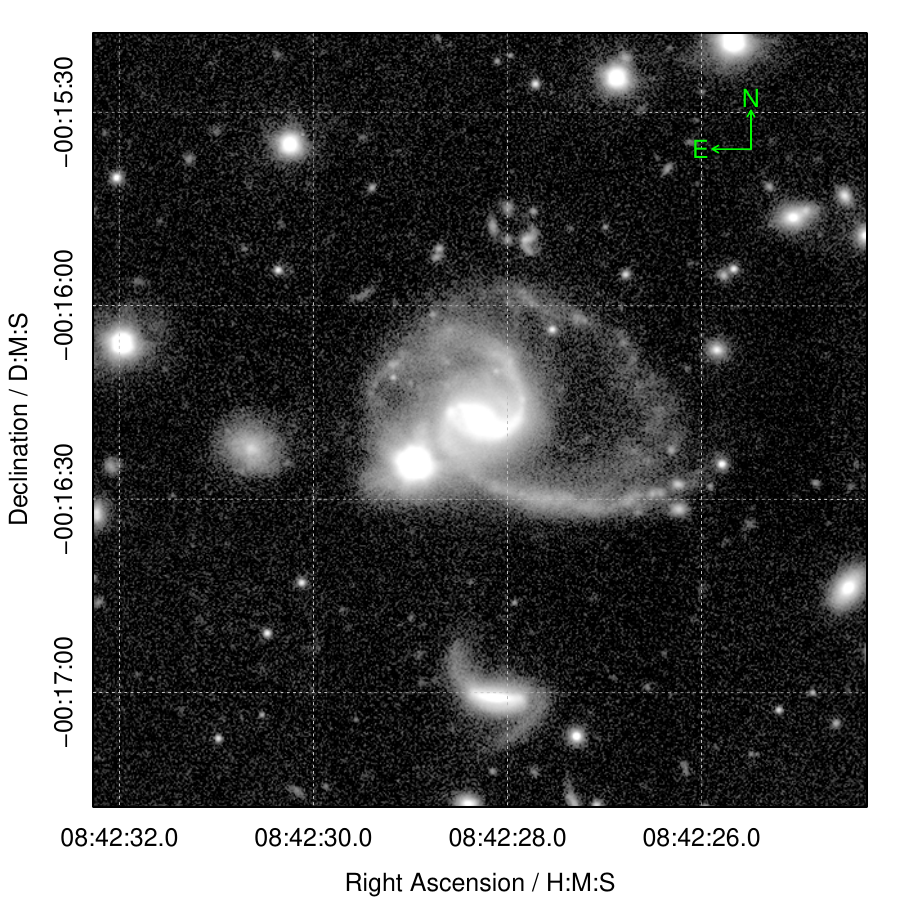}
 \caption{Example VST $r$-band image of a stored \FITS{} file loaded with \Rfits{} and displayed with the default plotting method. This uses \Rwcs{} to correctly display celestial coordinate axes.}
 \label{fig:example_wcs}
\end{figure}

There are a number of important tasks carried out by \Rwcs, but generally we use it to carry out one of two operations: convert image pixel positions to celestial coordinate positions, or vice versa. To carry out this conversion information is needed about the WCS of the image saved on disk, and this is provided in a relatively standardised manner via a number of key words stored in the \FITS{} header that describe key aspects of how the image is projected. This is why \Rwcs{} is so closely entwined with the \Rfits{} package. For flexibility, it is possible to use \Rwcs{} more directly (specifying values for relevant keywords in a laborious manner), but to minimise errors this approach is not recommended (and almost never used in practice). An example of the \Rfits{} and \Rwcs{} packages naturally interfacing is shown in Figure \ref{fig:example_wcs}, where we load the target \FITS{} file with \Rfits{} including all the relevant information regarding the WCS, the celestial coordinate axes are then overlaid using the \Rwcs{} package when we use the default plotting method on the in-memory \FITS{} image data.

\subsubsection{Supported Projection Schemes}

The are a lot of possibilities when it comes to representing a WCS in terms of axis types, projection, and the reference system used (these are presented as tables in Appendix B). Table \ref{tab:axis_type} shows the range of recognised axis types in \Rwcs, and by extension \propane{} (note this represents all options supported in \wcslib{} as of v7.9). In practice the majority of modern survey data uses Tangent Gnomonic (TAN) projection, with some popularity for Orthographic projection in the radio community (both SIN and NCP). In both cases this is because the telescopes in question natively produce images with close to this type of projection (bar correctable distortion terms). In the case of Tangent Gnomonic, all great circles are preserved as straight lines, but it does not maintain other aspects of equal area or shape. This can cause image artefacts in the corners of large (many degree) fields, so typically contiguous regions are never more than a couple of degrees across.

\subsubsection{Supported Distortion Schemes}

To achieve the warping of pixels from one WCS to another, correctly accounting for all aspects of image distortion is critical. Importantly, \wcslib{} (and by association \Rwcs) has been updated to include all the most recent paradigms of image distortion, where very high order polynomials (often more than $5^{th}$) are typically used to describe deviations from the analytic projections described above. Supported distortion schemes are: PV, SIP, TPV, DSS, WAT and TPD \citep[see][for a detailed discussion]{2002AandA...395.1077C}, where only the PV type polynomial distortions can be entered manually (the others require access to the full \FITS{} header).

This distortion flexibility in \wcslib{} has proven to be particularly important for projecting and warping James Webb Space Telescope (JWST) images, which use more modern SIP distortion terms than available in the standard public version of the popular \swarp{} warping and stacking software suite \citep{2010ascl.soft10068B}\footnote{www.astromatic.net/software/swarp/}. Some care has to be taken when warping images that have distortion terms since they will create local covariance features (pixels being pushed together) and are sometimes not even analytically invertible. For this reason, the target WCS that images are being warped onto should almost never contain distortion terms itself, but instead should be a fairly benign (distortion free) projection. Distortion terms can also create small geometric differences between forwards and backwards projection solutions because not all distortion schemes can be analytically inverted.

\subsection{imager}

To accomplish image manipulation operations, a number of low-level libraries exist. Basic tasks like image translation (integer pixel and sub pixel), rotation and rescaling have been worked on as computer science and software engineering problems for many decades. Anything from sprite manipulation in video games to photograph processing requires such routines, and this makes them amongst the most optimised tasks in modern computing \citep{citeCoreAlgo}. For this reason, rather than attempt to re-implement such manipulation from first principles, it is far more sensible to make use of an open-source library. For our purposes, we use the \imager{} package that interfaces with the \Cpp{} \CImg{} library of optimised light-weight image manipulation routines \citep{citeCImg}\footnote{cimg.eu}. \CImg{} v2.9.1 is included with the current version of the \imager{} package, so no alternative downloading or linking to this library is necessary.

\imager{} has existed as a public GitHub repository since May 2015\footnote{github.com/asgr/imager}, where it was originally developed by software engineers based in France. The current release is v0.45.5, and it is an extremely popular package within the \R{} ecosystem, with much of its usage coming from outside of astronomy (particularly biological and medical sciences). In early 2023 ASGR and RT took over as project maintainers. \imager{} is comprehensively documented with a 134 page feature-complete manual, passes all current CRAN package consistency checks, and has a large number of online examples and vignettes covering the majority of its capabilities. As such, we only briefly mention the key features relevant for \propane{} here. 

A huge array of image manipulation and filtering tasks are possible with \CImg{} and by association \imager, however the key routines of interest for \propane{} are its variety of image warping, interpolation, and stacking / combining routines. These are highly optimised \Cpp{} routines, with many allowing for parallel threading on central processing units (CPUs). In terms of image warping and interpolation (including re-gridding), the routines offered allow for:

\begin{enumerate}
\item bilinear
\item bicubic
\item Lanczos
\item nearest-neighbour
\end{enumerate}

interpolation, with each able to use forward or backwards projection \citep[see][for discussion on interpolation techniques]{citeCoreAlgo}. For most purposes bicubic or Lanczos approaches are the most accurate (in terms of features and / or flux conservation), with a small cost in speed compared to bilinear options. We note that the Lanczos algorithm is only available for on-axis re-sampling in the \CImg{} library (not general purpose warping).

The core mechanic that allows \propane{} to use these general purpose \imager{} routines is its method for describing the warp field of an image. In simple terms, \imager{} allows a user to create a warp field image for a particular pair of images (either forwards from the input image to the output, or vice versa). To create this warp field image requires two calls to \Rwcs{}. In the case of forward projection, the first call maps all pixels in an input image to celestial coordinates, and the second call converts these celestial coordinates back into pixel positions on the output image. These two operations mean we have a full mapping in pixel space between the two images. \imager{} then uses its built-in interpolation and re-mapping methods to distribute the flux in the input image to the target WCS, ensuring (as far as is possible) the accurate conservation of flux and features. There is always a tradeoff in the process however \citep{citeCoreAlgo}, so flux critical tasks (at the level of sub 1\% accuracy) should usually avoid any projection of the data entirely and work in native image space instead.

\section{ProPane}
\label{sec:propane}

\propane{} builds on the foundations of the lower-level packages discussed above to provide fully WCS aware image manipulation routines. Most fundamentally this includes the singular warping of an input image onto a target WCS (which is often useful for image alignment and allows for simple multi-band photometry with e.g.\ \profound), but by natural extension this includes the bulk warping of a large number of input images (in practice limitless) and later combining to create a stacked mosaic which may cover a much larger area than the original input frames and / or be much deeper in terms of signal-to-noise. This form of a large stacked mosaic is often the end-point for large campaigns of imaging that repeatedly target the same region of sky e.g.\ GAMA and COSMOS (see Table \ref{tab:acronym} for survey descriptions).

\propane{} has existed as a public GitHub repository since February 2023 \citep{ProPane}\footnote{github.com/asgr/ProPane}. It is used for a number of image warping, mosaicking and combining tasks by the astronomy community in its current v1.6 form, and considered largely stable in terms of features and performance. It is comprehensively documented with a 35 page feature-complete manual, passes all current CRAN package consistency checks, and has a large number of online examples and vignettes covering the majority of its capabilities\footnote{rpubs.com/asgr}.

\subsection{Image Warping}
\label{sec:image_warping}

We briefly discussed the warp field method above that allows \propane{} to combine \Rfits, \Rwcs{} and \imager{} to describe the pixel-to-pixel warping required to correctly translate the pixel grid in one image to the pixel grid in another. With possession of this warp field, there are still some decisions that must be made about how to best re-project the pixels as desired.

\begin{figure}
 \includegraphics[width=\columnwidth]{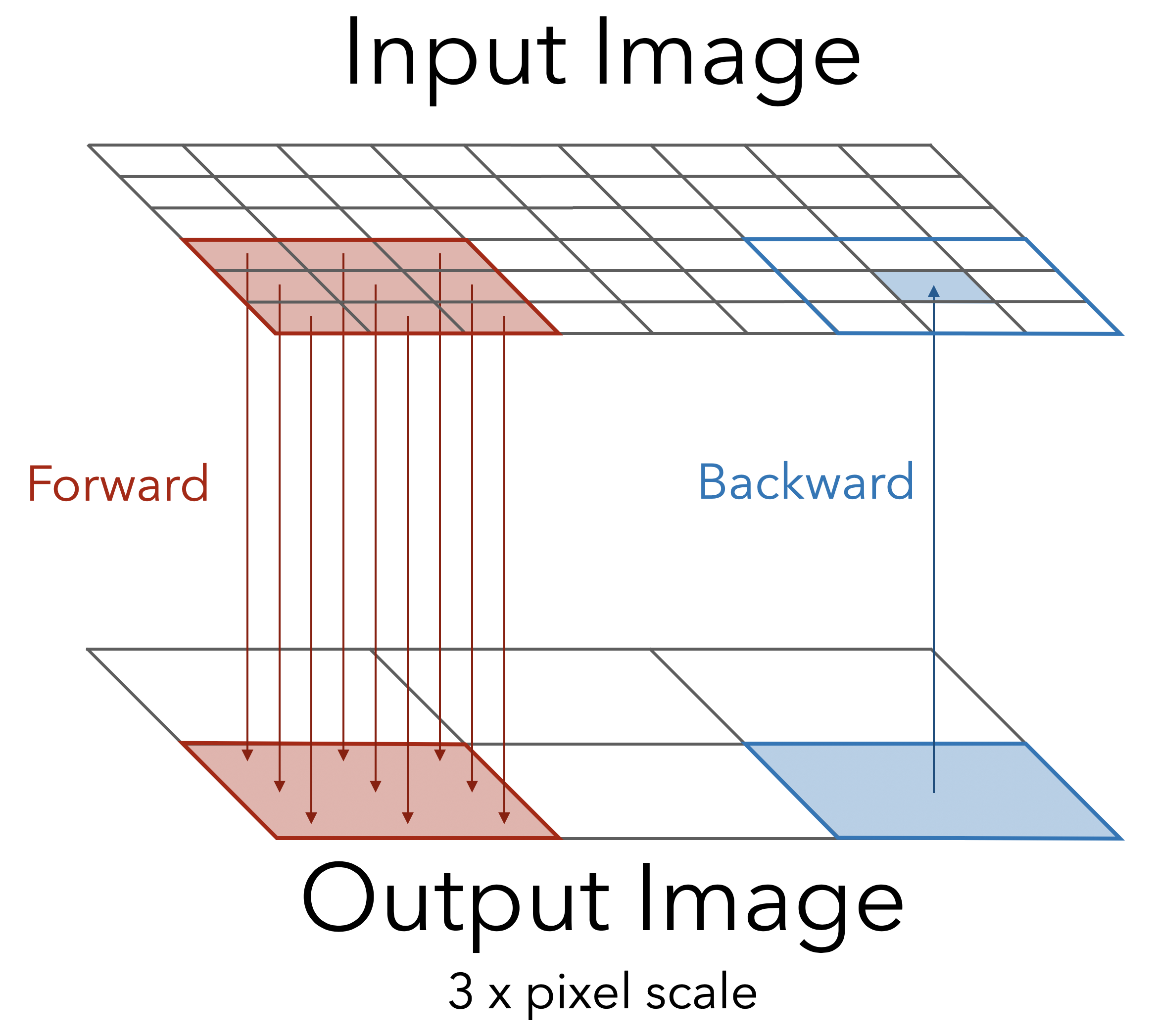}
 \caption{Schematic of forward (red) and backward (blue) image projection.}
 \label{fig:propagation}
\end{figure}

The tradeoffs when projecting image pixels are usually a combination of flux conservation, feature preservation (sharp discontinuities in flux are visually preserved) and computational cost \citep{citeCoreAlgo}. A basic decision when warping is to decide whether it is best to project all pixels in an input image forwards onto a target WCS (or image), or better to look at all output pixels and reverse the flux path backwards to find where on the input image the flux originates. Figure \ref{fig:propagation} is a simple schematic cartoon of how forward and backward pixel propagation differ when the input and output pixel scales vary by a factor of 3 (with the output image having coarser pixels). To test the impact of a target WCS on the pixel covariance, \propane{} includes the propaneCovTest function that projects a checkerboard pattern through the specified warping, allowing users to inspect different schemes objectively.

\begin{figure*}
 \includegraphics[width=\columnwidth]{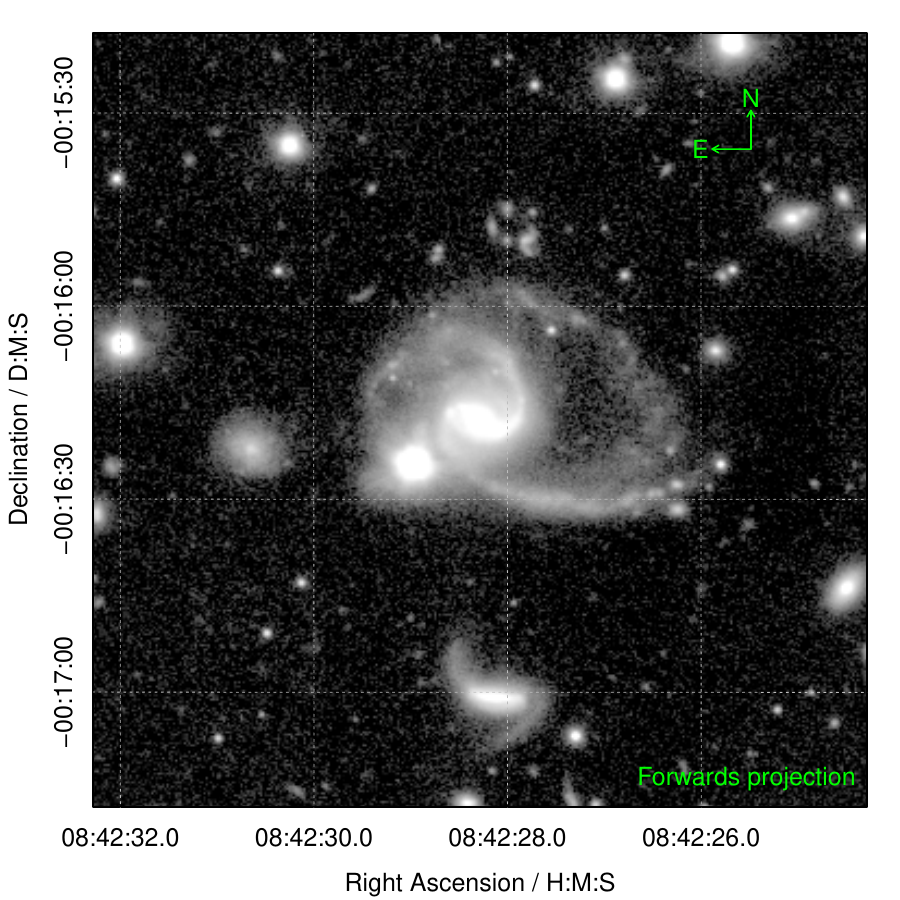}
 \includegraphics[width=\columnwidth]{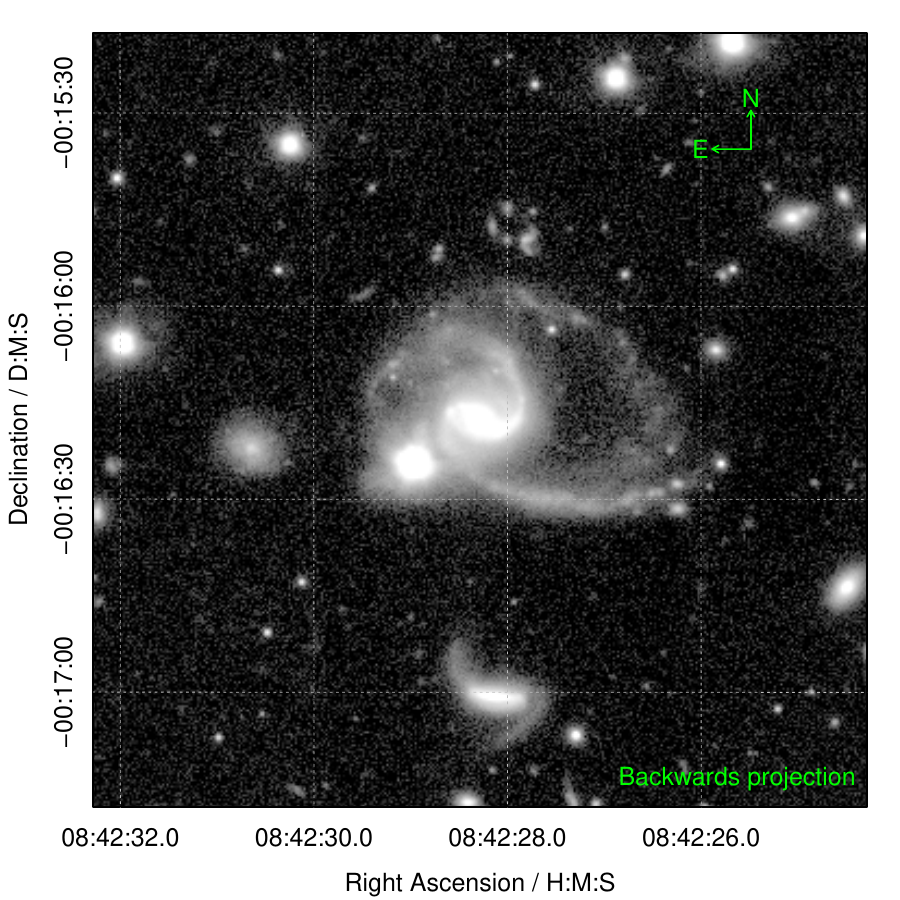} \\
  \includegraphics[width=\columnwidth]{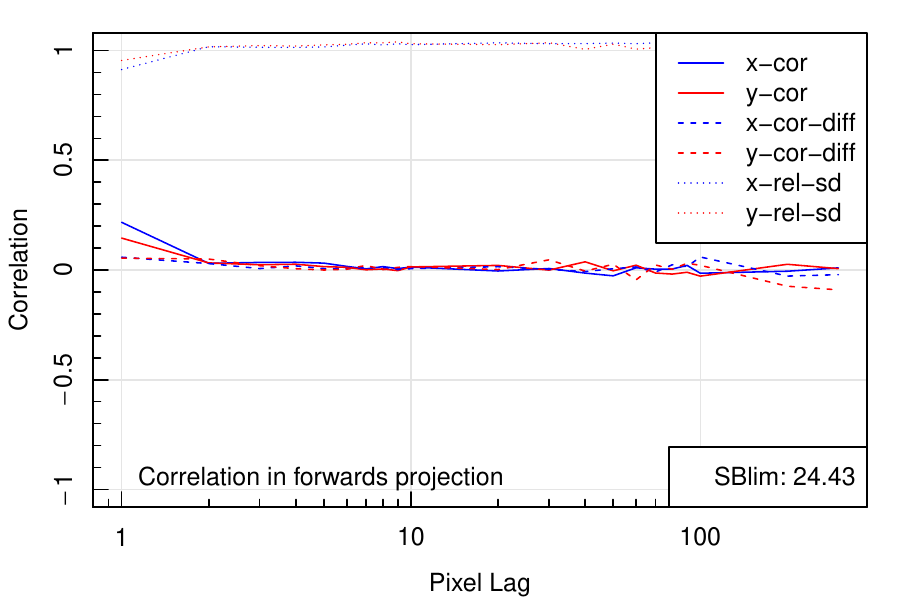}
 \includegraphics[width=\columnwidth]{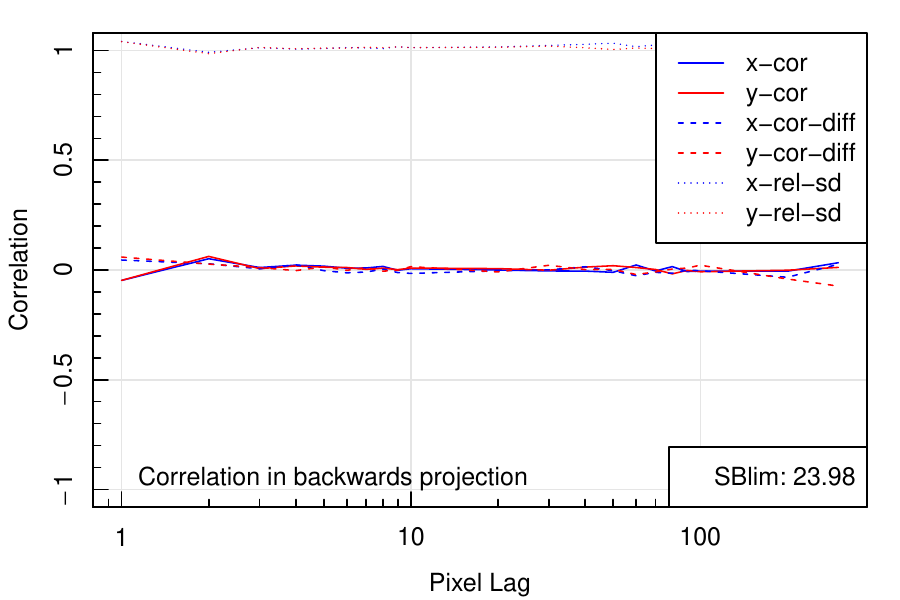}
 \caption{Comparison of depth differences (top) and pixel covariance (bottom) between forward projected pixels (left) and back projects pixels (right). In both cases a VST image is being projected onto a VISTA image WCS. For the covariance plots, the solid line is the $x$/$y$ (blue/red) pixel lagged covariance (nearer to 0 is better), the dashed is the difference between the covariance of positive and negative sky regions (nearer to 0 is better); and the dotted is the root mean square (RMS) normalised standard deviation between lagged pixels (nearer to 1 is better).  The SBlim is the $5\sigma$ surface brightness depth of the image in AB mag. Whilst forward projection creates a 0.45 mag deeper image, it suffers more from pixel covariance at the 1 pixel lag scale.}
 \label{fig:for_v_back}
\end{figure*}

Figure \ref{fig:for_v_back} is an example of both directions of warping, where we see a Very large telescope Survey Telescope (VST) $r$-band image projected onto a target Visible and Infrared Survey Telescope for Astronomy (VISTA) K-band WCS. Generally speaking, forward projection \citep[more akin to drizzling, see][]{2002PASP..114..144F} will preserve flux better and produce deeper images, but backward projection produces fewer image artefacts because the sampling rate in the output image is consistent throughout the WCS. For related reasons, forward projection tends to produce more obvious signs of pixel image covariance and generally slightly better signal-to-noise across the image. These behaviours are clear in Figure \ref{fig:for_v_back}, where we find the forward projected image is 0.45 mag deeper than the backwards projected image. The difference in pixel scale in this example is $0.2\arcsec$/pix (VST) versus $0.34\arcsec$/pix (VISTA), which would suggest an upper limit to the increased depth of 0.58 mag. In practice, projection and interpolation kernel artefacts (that also cause the covariance described) prevent this full depth advantage from being realised.

Closely related to the concept of pixel covariance is sub-pixel sampling to increase the effective resolution of an output image. The principle is that sub-pixel shifts in input images might allow higher resolution outputs, a process generally referred to as dithering \citep[e.g.\ see][]{2011ApJ...741...46R}. To what degree this increase in effective resolution is possible is a strong function of the amount of sub-pixel shifts present, and the algorithm used to warp and stack images. We ran tests of approximately critically sampled images generated with \profit{} \citep{2017MNRAS.466.1513R}, where the input Moffat PSF has a three pixel full-width half-maximum (FWHM) profile and a concentration of 2. The target of the test was to double the effective resolution via image dithering (creating sub-pixel offset input frames), followed by resampling (to increase resolution) and warping (to re-align the frames). The Lanczos algorithm for image re-sampling (available in \CImg) allows for accurate feature preservation (in terms of the effective FWHM when up-sampling), with only very minor improvement when increasing from coarse half sub-pixel ($2 \times 2$) to fine quarter sub-pixel ($3 \times 3$) dithers. The final measured FWHM was effectively identical to a reference image created natively with a FWHM of 6 pixels, but there was about 7\% flux loss in the brightest pixel of the PSF in both dither variants (where more than 99.999\% of the flux lost is spread into adjacent pixels). The main consequence of under-sampling an image and dithering/warping with \propane{} would likely be flux preservation (as discussed elsewhere in this work) rather than feature preservation. In this regime a pure drizzling approach might be preferred \citep{2002PASP..114..144F}.

\begin{figure}
 \includegraphics[width=\columnwidth]{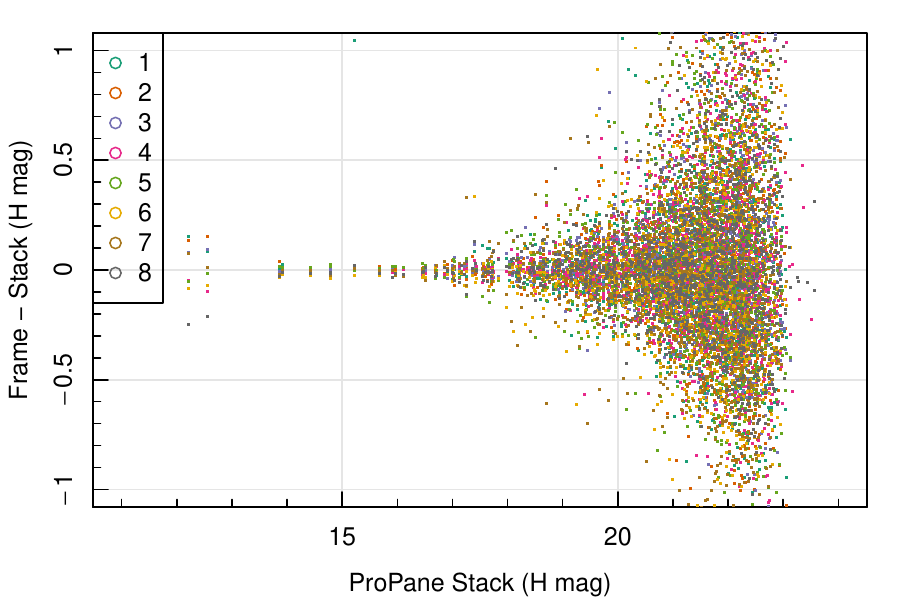}
 \caption{A comparison of the source magnitudes for eight native resolution images compared to the magnitude of the final warped and stacked source (these data are more fully discussed in Section \ref{sec:image_prepare}).}
 \label{fig:input_frames_mag}
\end{figure}

To further explore the impact of warping objects with the default scheme available in \propane (via \imager{} and \CImg), Figure \ref{fig:input_frames_mag} shows the scatter between eight input frames and the final mean stacked image that are discussed in more detail in Section \ref{sec:image_prepare}. The average $1\sigma$ scatter between input frame and stacked magnitudes for stars brighter than $17^{th}$ mag is 0.015 mag, largely reflective of zero point error. Fainter than this limit the scatter is dominated by sky subtraction variations. The two brightest stars (near to $12^{th}$ mag) have more scatter than those immediately fainter--- this is a consequence of the individual frames having saturated pixels and spurious photometry as a consequence. In general, there is a small amount of flux loss induced by the warping process and the underlying interpolation scheme employed by \CImg{}. For bright stars (i.e.\ compact sources) this is $\sim0.01$ mag, and less for more extended sources. This is broadly consistent with later comparisons made to \swarp{} in Section \ref{sec:image_stacking}.

The CPU time and memory required to warp an image scales fairly directly with the product of input and output image pixels. In the example above a $602 \times 602$ VST image being warped onto a $356 \times 356$ VISTA target WCS takes approximately 0.25s on a relatively basic 2017 MacBook using a single core. 60\% of the time is spent creating the warp field via \Rwcs; 20\% projecting the image via \imager; and 20\% other tasks (e.g.\ restructuring the data to create an \Rfits{} object output). If repeated warping operations are going to be carried out then it is possible to save the warp field information, speeding up the process by more than a factor of two typically.

\subsection{Segmentation Map Warping}

A special use case for image warping is projecting segmentation maps and masks to a different WCS. In these cases you generally do not want to conserve any pixel intensity, and instead want to preserve the integer nature of the input segmentation map. \propane{} allows users to do this via the propaneSegimWarp function, which means it is viable to project segmentation maps between different WCS.

\begin{figure*}
 \includegraphics[width=\columnwidth]{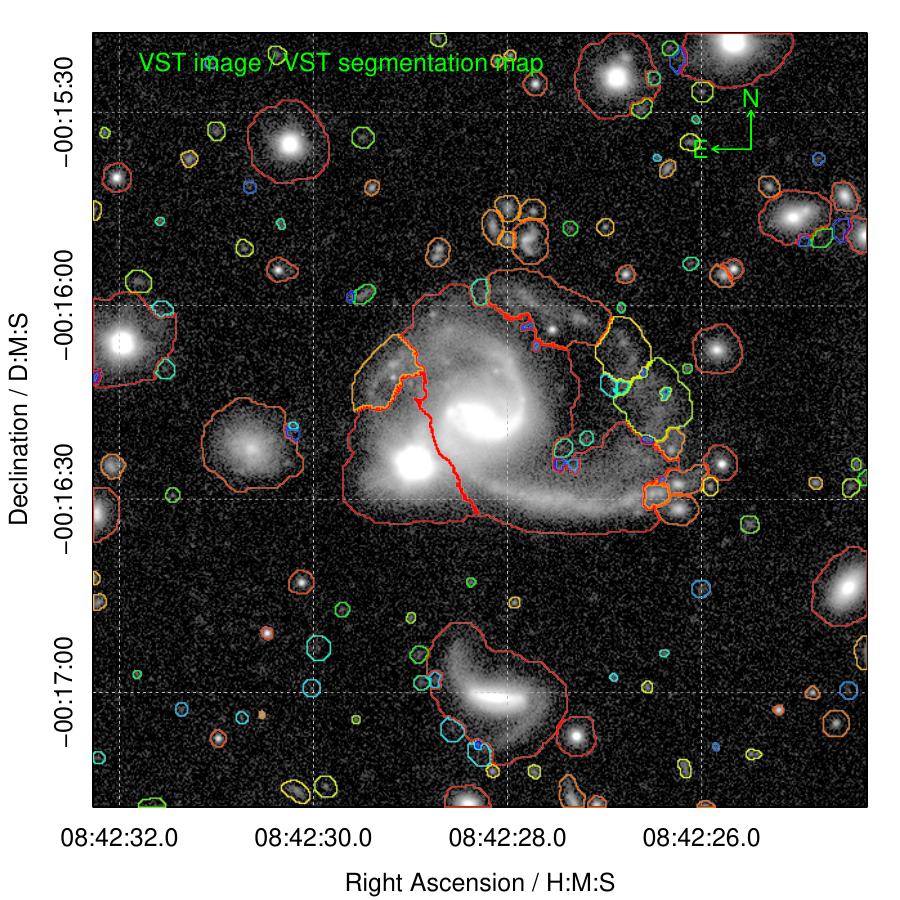}
 \includegraphics[width=\columnwidth]{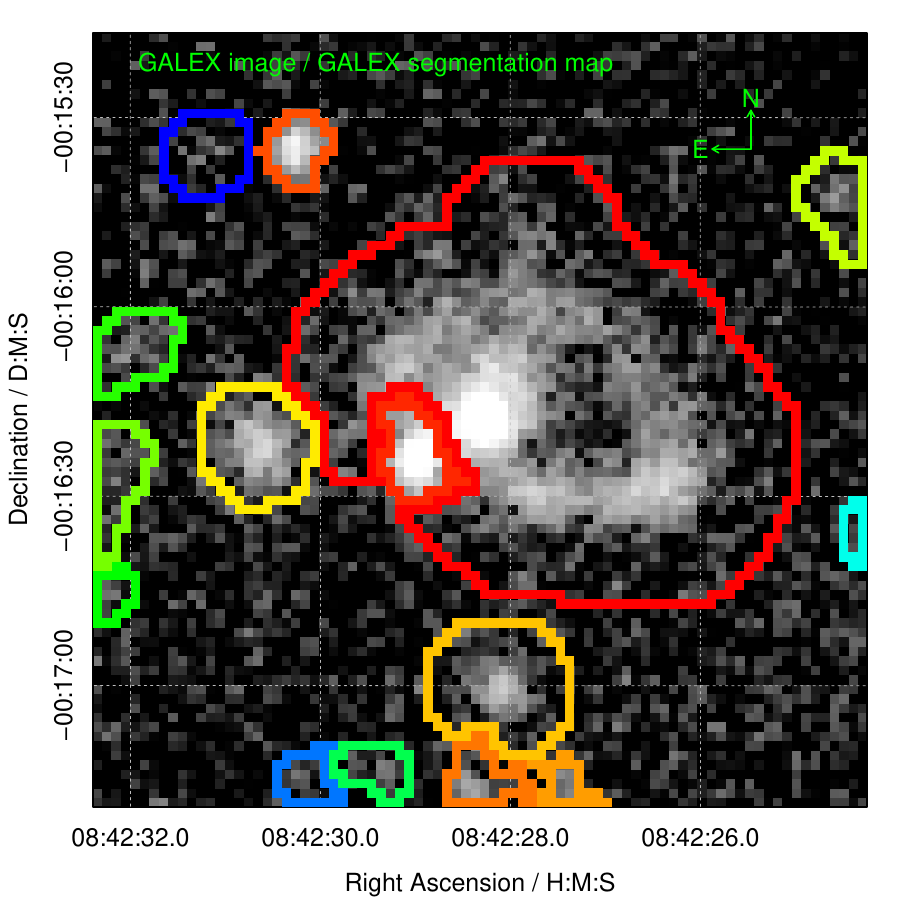} \\
  \includegraphics[width=\columnwidth]{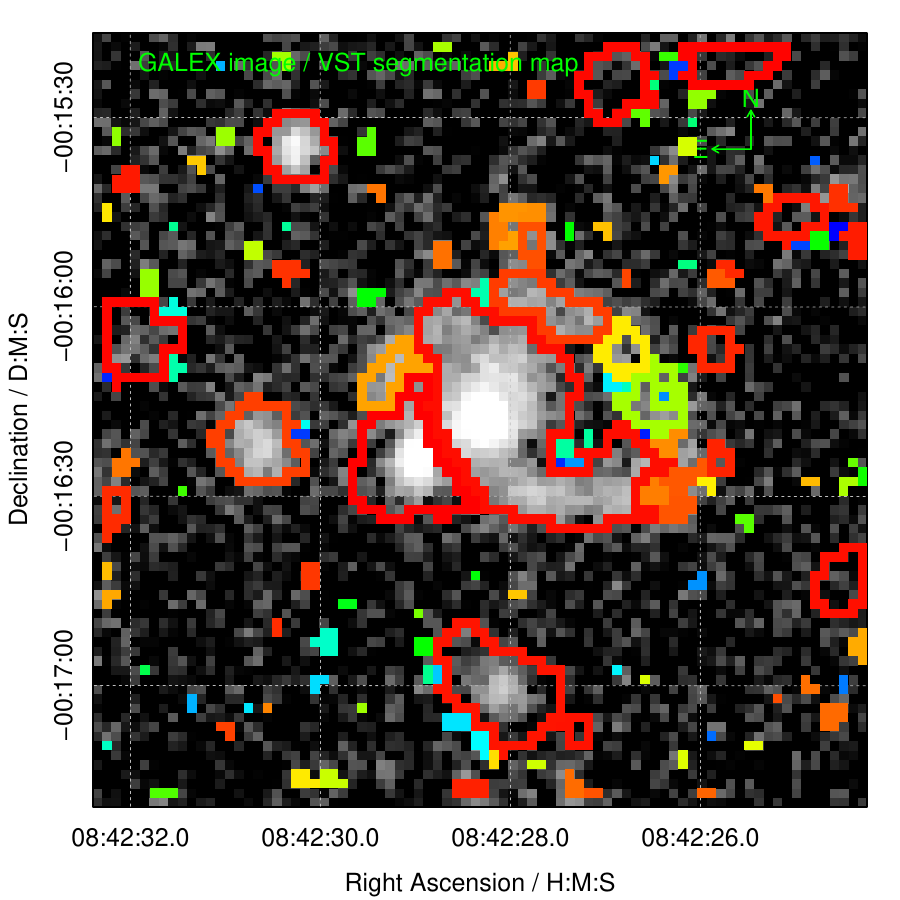}
 \includegraphics[width=\columnwidth]{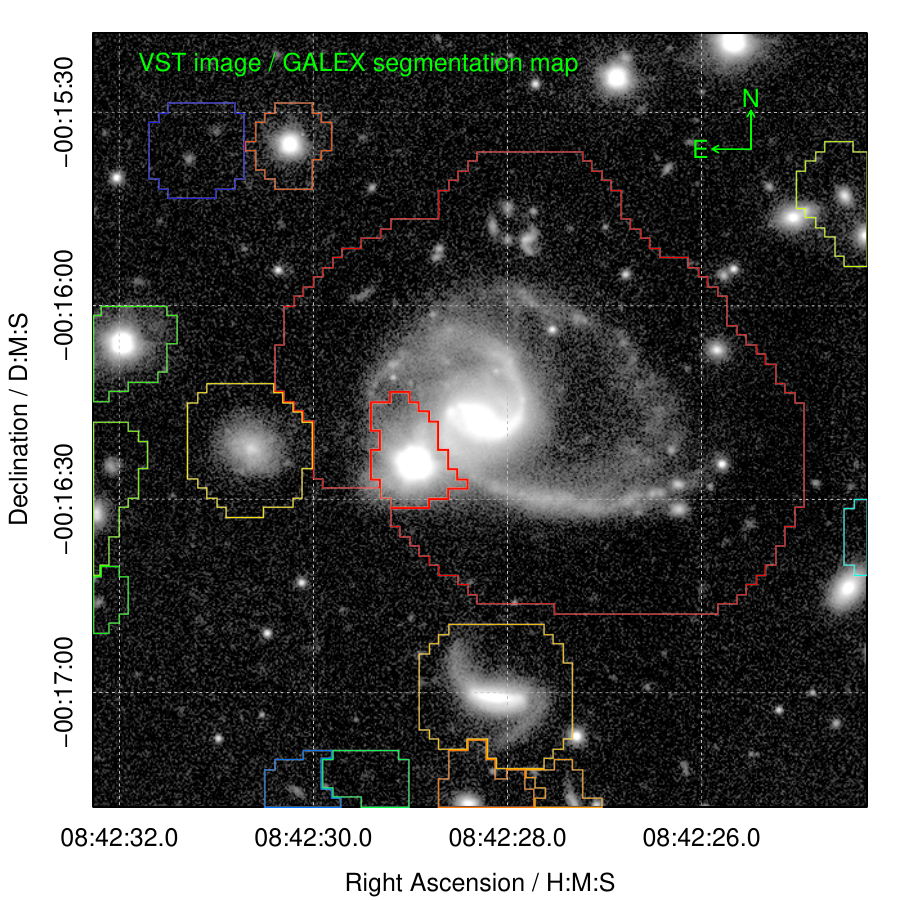}
 \caption{Top panels show the native segmentation maps generated with \profound{} for VST (left) and GALEX (right) data. Bottom panels show the warped VST segmentation map overlaid GALEX (left) and the GALEX segmentation map overlaid on VST (right).}
 \label{fig:segim_warp}
\end{figure*}

In the situation where you have a mixture of input photometry with very different pixel scales, warping and re-projecting the segmentation map is often the better solution when trying to do forced photometry because you will avoid the introduction of pixel covariance and small flux conservation artefacts (as discussed above). This segmentation warping approach has been used in \citet{2022MNRAS.516..942C} which required the preparation of multi-band segmentation maps for galaxy profiling with \profit{} \citep{2017MNRAS.466.1513R}.

Figure \ref{fig:segim_warp} shows an example of native VST and Galaxy Evolutionary Explorer (GALEX; an ultra violet space telescope) segmentation maps warped onto the other WCS. The obvious consequence of this is some sources that are clearly resolved in VST cannot be separated with GALEX data, but it does ensure that the photometry is appropriately matched whilst avoiding the introduction of image artefacts caused by warping the image pixels instead. There are also some segments in the VST data that exist fully within a GALEX pixel, and are not represented at all after warping (so the resulting catalogue will not necessarily have GALEX photometry for every VST object).

\subsection{Image Stacking Preparation}
\label{sec:image_prepare}

Conceptually, image combining means taking a collection of images, and determining properties on a common grid of celestial coordinates. A popular use case might be to warp multiple images onto a common WCS, and then simply calculate the weighted mean of flux on all pixels. This type of image combining is usually called `stacking', but there are a few competing methods to achieve optimal outcomes even in such a simple scenario.

The common problem to solve for all image combining methods is determining which frames overlap within a given region of sky, and following that what type of statistics should be computed. The following subsections describe how these problems are solved in \propane.

\subsubsection{Finding Overlapping Frames}

It will often be the case that we have already curated a list of images that we know to significantly overlap on sky, but with modern large surveys we often have datasets of thousands of images that need to be assessed for spatial overlap. \propane{} has a bespoke function (propaneFrameFinder) to facilitate the efficient discovery of overlapping data. In brief, this searches for all frames that even partially overlap with a given position and search radius. It does this only inspecting the header component of all target \FITS{} files, so there is very little computational cost to scanning very large directories of images (and this process can even be run in a multi-core manner).

\begin{figure}
 \includegraphics[width=\columnwidth]{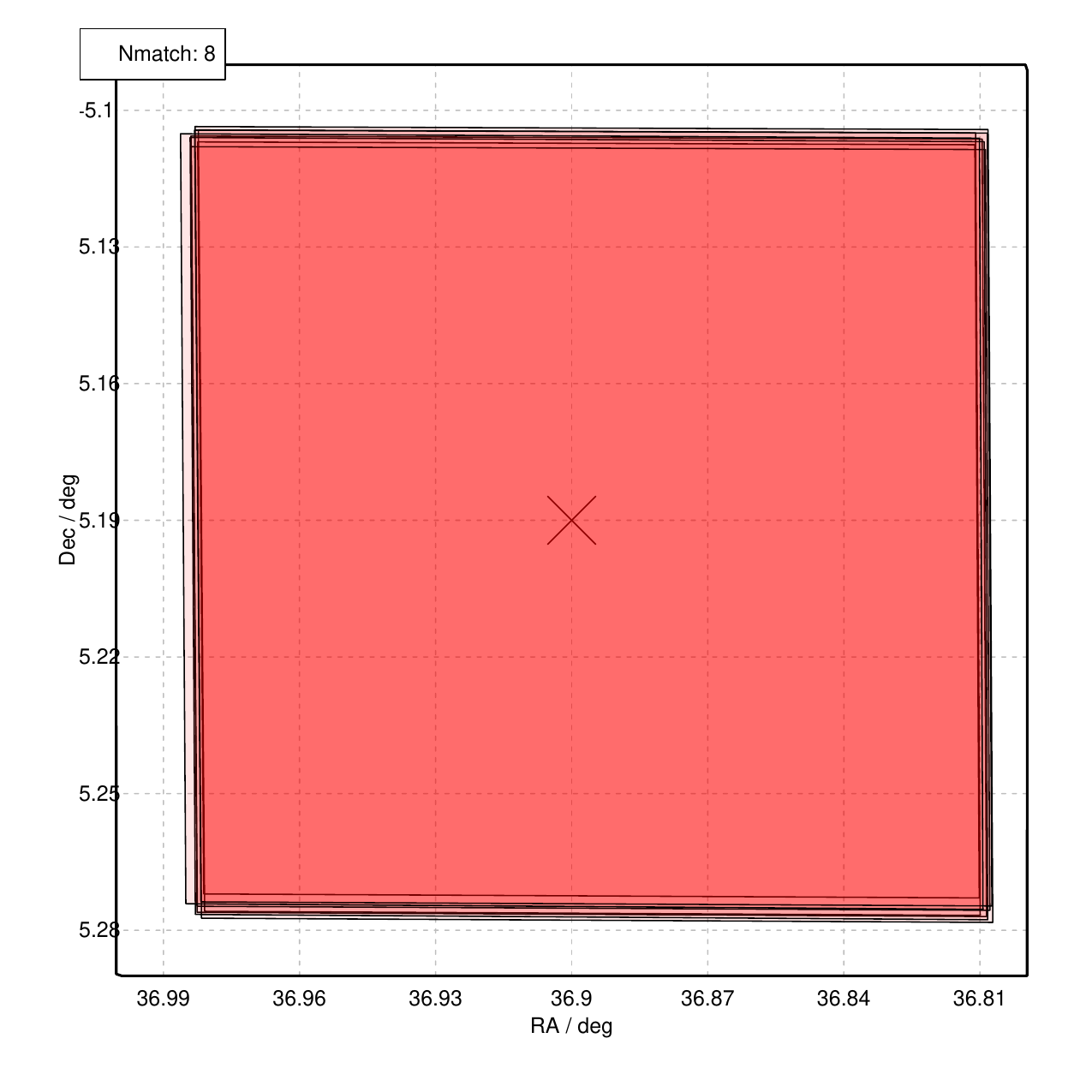}
 \caption{Example of propaneFrameFinder detecting all VISTA VIKING frames that overlap within $6'$ of a target position (shown as a cross). Here it recovers eight overlapping frames.}
 \label{fig:frame_find}
\end{figure}

Figure \ref{fig:frame_find} shows the overlapping VISTA Kilo-degree Infrared Galaxy survey \citep[VIKING;][]{2013Msngr.154...32E} H-band frames for an example run searching for frames overlapping within $6'$ of 36.9 deg (RA) and -5.19 deg (Dec). The plot shown is an optional output of propaneFrameFinder, where the geometry of overlapping frames might inform a different target WCS being chosen to project the frames onto.

\subsection{Image Alignment}

\begin{figure*}
 \includegraphics[width=18cm]{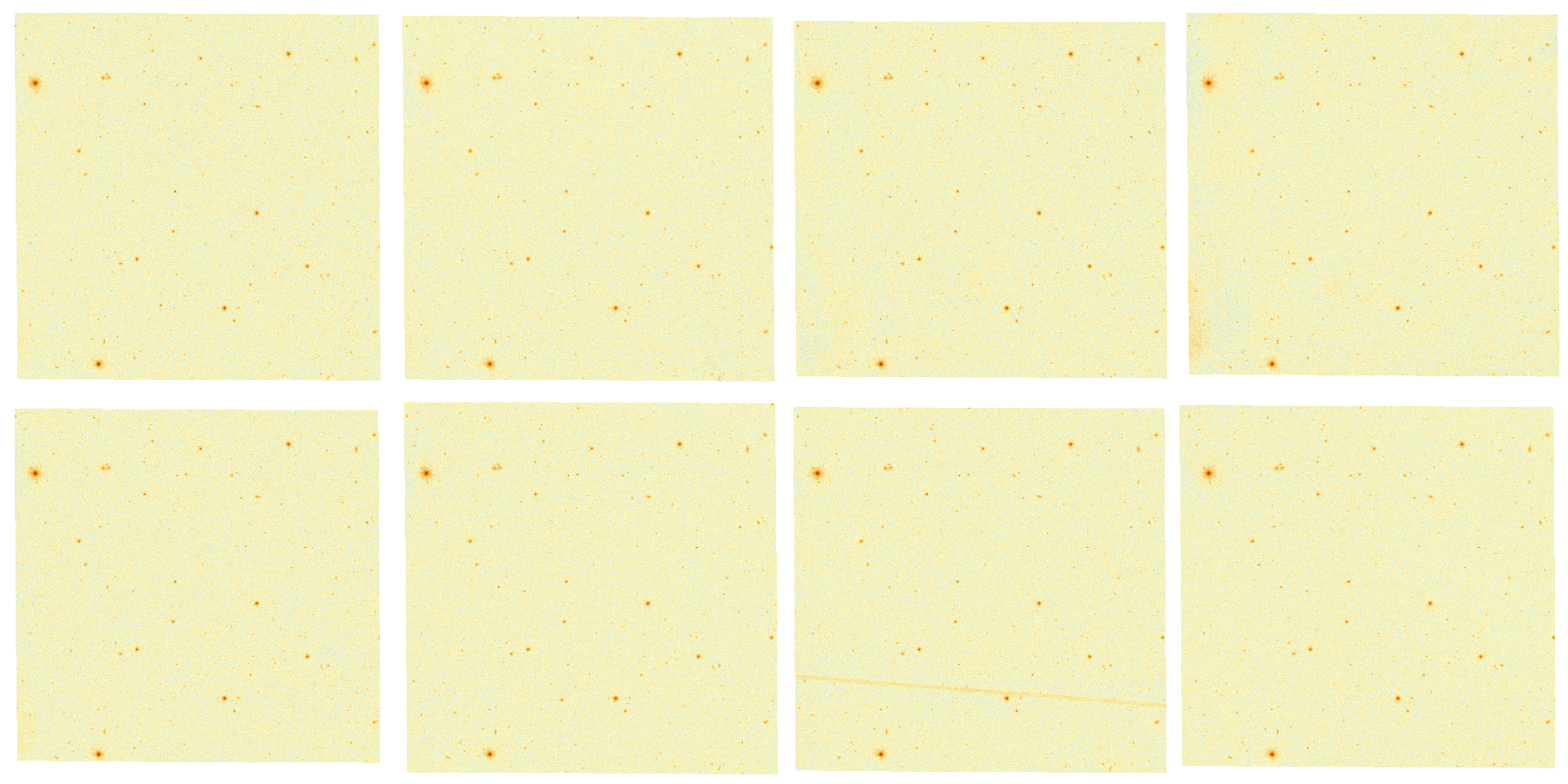}
 \caption{The eight overlapping VISTA frames presented in Figure \ref{fig:frame_find}, but now warped to a common WCS. Redder pixels are positive flux and bluer pixels are negative flux (yellow being zero).}
 \label{fig:input_frames}
\end{figure*}

Once the overlapping frames have been identified users tend to be in one of two paradigms: either the WCS of frames overlap but the pixels are not aligned (which will necessitate image warping / projection); or the pixels are exactly aligned already. In both of these paradigms we are assuming that the input images in question have already been properly background subtracted using tools such as \profound{} (in our case), so the background level sits numerically at zero. If images need to be aligned then the high-level propaneStackWarp functions will project all images to a common target WCS using propaneWarp (with access to all the options discussed above). Figure \ref{fig:input_frames} shows the eight overlapping images found in Figure \ref{fig:frame_find} but now warped to a common WCS.

\subsubsection{Tweaking}

A common problem when trying to stack and combine astronomical images is that the WCS (though usually specified in the header) might be inaccurate. This might be because of issues with the guide stars used to define the WCS, i.e.\ there might not be many present in a small field-of-view, or they have large proper motions. It can also be caused by algorithmic errors when processing and reducing the images, i.e.\ the fitted WCS solution might be poorly converged because of low signal-to-noise in the image etc. To combat this problem it is often necessary to re-align the images in question and adjust either the image pixels or the WCS contained in the header. This process is commonly known as `tweaking'. Many science cases (basic multi-band photometry) only require images to be well aligned at the pixel scale, but some lensing science requires even sub-pixel tweaking. Small misalignment can be thought of as a form of artificial blurring when stacking. In general, the better aligned the input images when stacking the deeper the final stacked image, so even if the resulting stacks are not obviously in error (e.g.\ double sources present in regions of the image) tweaking might be beneficial.

Because of the small field-of-view of Hubble Space Telescope (HST) / James Webb Space Telescope (JWST) and astrometric errors in guide star positions (often due to the proper motions of stars), a large fraction of the data needs some degree of tweaking to allow for well aligned stacking and deep multi-band photometry. For our JWST PEARLS\footnote{See Table \ref{tab:acronym} for survey description.} work in particular, the pixel and catalogue-based tweaking discussed below have allowed us to correctly align large numbers of images for stacking purposes. This allows for the generation of accurate forced photometry catalogues using \profound, and the subsequent fitting of SEDs with \prospect.

As a general note, tweaking is possible between different images with different filters as long as they share the majority of source. E.g.\ stars tend to emit radiation across a large range of wavelengths, and will have the same intrinsic position on-sky in all cases. Where tweaking might break down is when the images come from hugely disparate wavelength regimes because very few sources are related, e.g.\ tweaking a radio continuum image to X-ray data (possibly no sources in an image emit at both wavelengths).

In the below all tweaking operations refer to strictly affine transformations that keep parallel lines parallel (i.e.\ scale, translation and rotation). More complex distortions are not supported by \propane{} currently, but other software exists that does support higher order terms \citep[e.g.\ SCAMP;][]{2006ASPC..351..112B}.

\subsubsection{Catalogue-Based Tweaking}

The simplest method of tweaking misaligned images is to match positions of sources in catalogue space. Usually this involves running a photometric extraction program (\profound{} in our case) and comparing astrometric positions between sources. Assuming the WCS errors are small (usually tweaking is at the scale of a few arc seconds) then it is usually simple to approximately match the {\it same} sources, and then adjust the astrometry for differences in translation and / or rotation (assuming the pixel scale is correct, but if it is not then you can also tweak for pixel scale). This process was carried out manually in \citet{2020MNRAS.496.3235B} in order to align GAMA catalogues with the latest Gaia reference catalogues \citep{2016AandA...595A...1G}.

\propane{} offers the propaneTweakCat function to allow such catalogue-based tweaking. To optimise the tweak solution an objective cost function has to be minimised (or maximised), and in the case of propaneTweakCat this is simply the sum of the squared residual position differences (which we wish to minimise, and would be zero if the catalogues perfectly aligned). Outlier rejection is offered as a user-defined parameter `delta\_max'. The actual optimisation is always done in pixel space, so the WCS of both catalogues has to be used to convert apparent celestial coordinates into $x$ / $y$ positions on a target WCS (usually that of one of the catalogues). Internally we use the quasi-Newtonian BFGS algorithm to optimise the solution \citep[Broyden, Fletcher, Goldfarb, and Shanno;][]{10.1093/comjnl/13.3.317}.

Catalogue-based tweaking is very fast, even when optimising for thousands of sources. The caveat is that object positions are not objective truths --- objects with complex geometry (say a merging galaxy) might have a truly ambiguous source position which might genuinely vary as a function of wavelength (since we might be tweaking across wavelength). Even when the source is simply a point source (e.g.\ a star) saturation and scattered light artefacts might still make it difficult to identify the source `centre'. Generally catalogue-based tweaking works best when the catalogues are trimmed of very extended sources (with potentially ambiguous centres) and of very bright objects with the potential for saturated central pixels. Also, fainter sources should generally be removed because they might be spurious, and even if they are real the lack of flux can cause errors in the positional determination. Generally the best results will be obtained when using the high $S/N$ catalogues that have no artefacts present, with good spatial coverage and source density throughout the image. Minimally five objects are needed to compute an unambiguous affine transformation, but more sources are preferable, especially to reduce the chance of spurious alignment (image sources being matched to erroneous similar configuration catalogue sources).

\subsubsection{Pixel-Based Tweaking}

More information exists in images than simply the central source positions used for catalogue tweaking. If pixels are directly compared between two images, generally the best astrometric solution is the one that minimises the apparent differences in projected flux. This will be the case even when the sources are not being observed in the same wavelength, since some flux overlapping on sky is still better (generally speaking) than none at all. There are more potential cost functions that could be optimised in the case of image based tweaking. We investigated whether it was more robust to minimise the product of two images, the pixel flux differences, or the square of the pixel flux differences. In general, the last objective function seemed to be the most stable and successful when comparing a variety of images.

Pixel-based tweaking clearly has the potential to use much more information and does not rely on decisions around photometric extraction and catalogue trimming. However, warping an entire image to a modified WCS will be an expensive process for large images. To mitigate this effect, it is possible to run the tweaking in `quick' mode, where only the brightest pixels are projected into a new WCS during tweaking (i.e. the others are ignored and assumed to be zero). This approach is generally orders of magnitude faster than warping an entire image, since only a few percent of pixels need to be transformed during each iteration of the optimisation. It is still necessarily much slower than catalogue-based tweaking though, since a typical bright source will still be hundreds of pixels (rather than a single position in a catalogue).

\subsubsection{Tweaking Comparison}

Generally if images have extremely large astrometric errors in the WCS, an initial round of catalogue-based tweaking is the best place to start because it can efficiently search a very large parameter space of translation and rotation. Once a reasonable solution has been found, a final round of pixel based tweaking can often serve to further improve the alignment solution. In terms of the output product, users can either chose to update the pixels in the image directly (leaving the nominal WCS in the header alone) or leave the pixels alone and update the WCS in the header. The latter is usually the better strategy if the resulting image will be used for further stacking, since you avoid creating unnecessary additional pixel covariance.

To compare tweaking performance, \propane{} comes with a small $356 \times 356$ test image. This is translated by +2 / +2 pixels in $x$ / $y$ and rotated $5^\circ$. A simple catalogue-based tweak is accurate to sub-pixel ($\sim0.5$ pixels) in translation and almost exactly correct with regards to the rotation and takes 0.04s. A full pixel-based tweak (warping all image pixels) returns almost exactly the right tweak solution and takes 46.81s, and the approximate pixel-based tweak (warping only bright pixels) is accurate to $\sim0.4$ pixels and takes 8.02s. Clearly catalogue tweaking is the fastest method, but also the least accurate. Pixel-based tweaking using the `quick' method is in practice a reasonable compromise between speed and accuracy for most purposes.

There are a number of options that allow fine control over how the tweaking operates, but the default parameters have been chosen to work well over a range of image types. Interested users should inspect the documentation and the extensive examples included in \propane{} to better optimise the behaviour of their tweaking.

As a general remark, catalogue-based tweaking is sensitive to the quality and depth of the reference catalogue. In the example above, this was generated using \profound{} on a reasonably shallow image with few high $S/N$ point sources. If a better quality reference catalogue is available (e.g.\ Gaia) then better quality catalogue tweaking will be possible. Indeed, JWST uses Gaia catalogue to tweak MAST pipeline products to sub pixel accuracy.

\subsubsection{Tweaking Example}

\begin{figure}
 \includegraphics[width=\columnwidth]{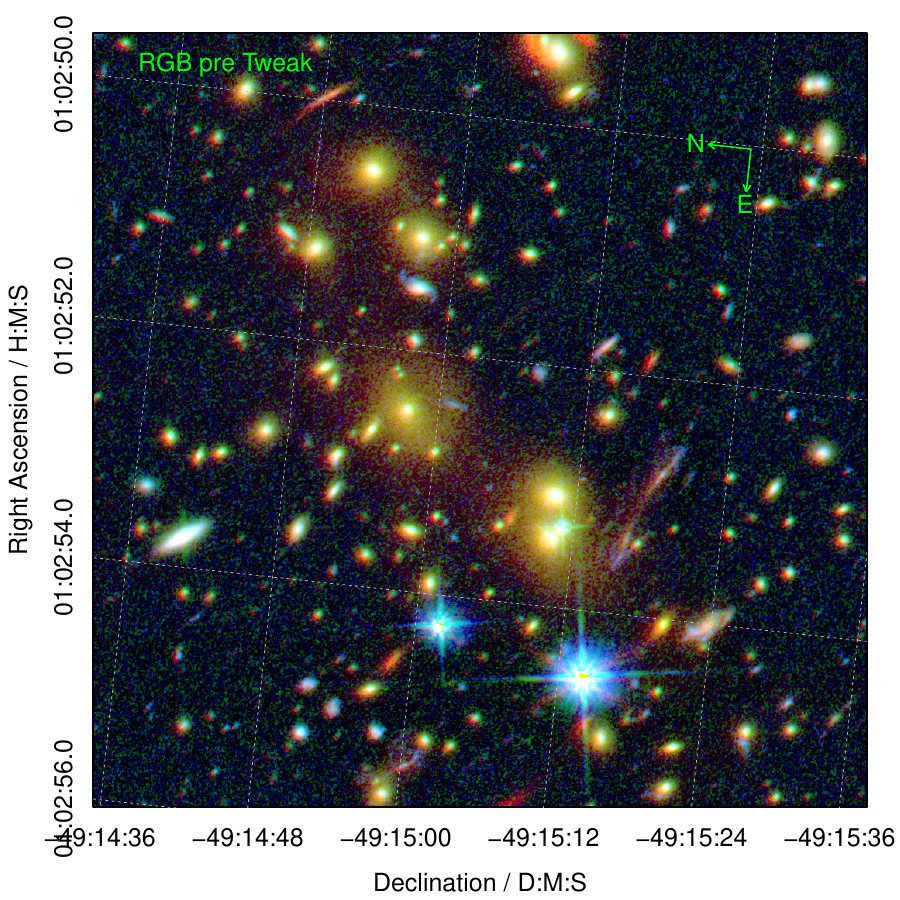}
 \includegraphics[width=\columnwidth]{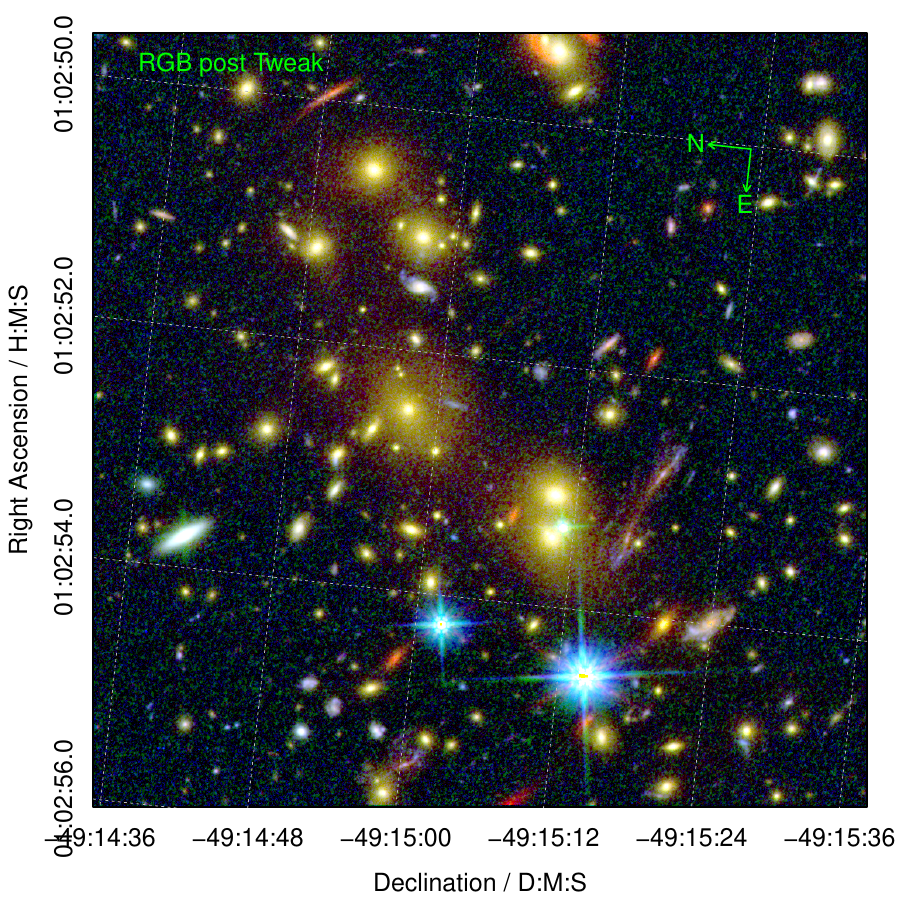}
 \caption{Top: RGB made pre tweaking, where the R is mapped to JWST NIRCam F444W, G is mapped to HST WFC3 F140W, and B is mapped to HST ACS F606W. Bottom: RGB post tweak, where the JWST image has now been correctly aligned with the HST data.}
 \label{fig:tweak_ex}
\end{figure}

An example of cross facility tweaking can be seen in Figure \ref{fig:tweak_ex}, where the JWST is misaligned by -5 pixels in $x$ and -4 pixels in $y$. Not accounting for this shift creates clear misalignment in the RGB image (where the R filter is mapped to JWST in this case). Running the frame through the pixel based tweaking (using the HST WFC3 frame as the reference image) creates a well aligned final RGB image, with the visually distracting colour ghosting entirely absent. This example $1000 \times 1000$ image took 13.8s to pixel align using default settings in the propaneTweakImage function.

\subsection{Image Stacking and Statistics}
\label{sec:image_stacking}

At this stage, we should be dealing with pixel-aligned images (although see the caveat regarding image tweaking below). At this point there are a number of ways we might want to combine and / or stack the data to reveal important information about the image statistics, e.g.\ for aligned pixels:

\begin{itemize}
\item {\bf sum} of all values per pixel
\item {\bf mean} of all values per pixel
\item {\bf weighted mean} of all values per pixel
\item {\bf median} of all values per pixel
\item {\bf quantile} of all values per pixel (i.e. this does not have to be the median, which is quantile 0.5)
\item {\bf maximum} of all values per pixel
\item {\bf minimum} of all values per pixel
\item {\bf standard deviation / variance} of all values per pixel
\end{itemize}

where in all the cases above we ignore pixels with missing information. In practice the two most popular image combination operations are a weighted mean (where the weighting is the inverse variance of the input images) and a median. Computing the inverse variance weighted stacked mean is achieved with the propaneStackInVar functions (with both a warping and flat / aligned stacking variant). This operation is trivial to parallelise since the weighted stack can be added to incrementally as long as the inverse variance stack is monitored alongside the combined image.

Considering first the inverse variance weighted stacking, \propane{} can use single values per image to represent the inverse variance, or potentially a full pixel map of inverse variance values. A map such as this might be the output of background analysis software (such as \profound), but also potentially the output of a previous \propane{} run (meaning it is easy to incrementally stack an image over time). The inverse variance map provided can capture purely the background RMS characteristics, or if a user desires the Poisson shot-noise from source flux can be incorporated by combining this in quadrature with the background RMS. In practice, capturing the smooth background RMS features alone is often preferable, especially since source shot-noise can be hard to estimate robustly.

\begin{figure*}
 \includegraphics[width=\columnwidth]{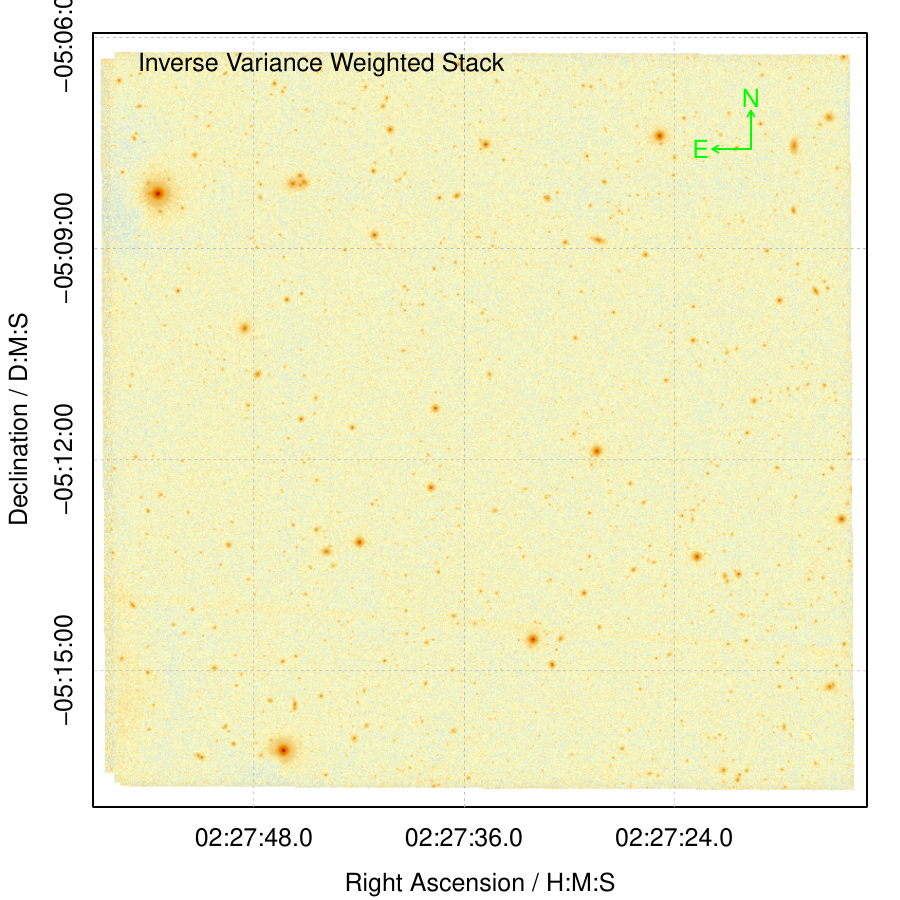}
 \includegraphics[width=\columnwidth]{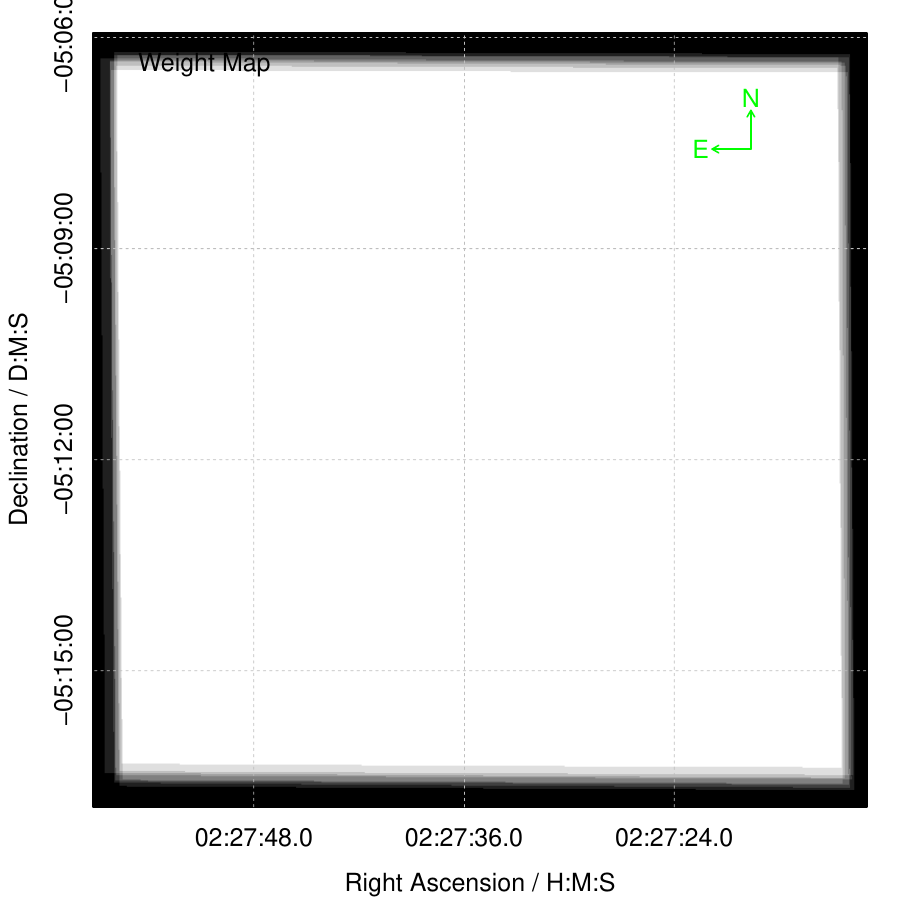} \\
 \includegraphics[width=\columnwidth]{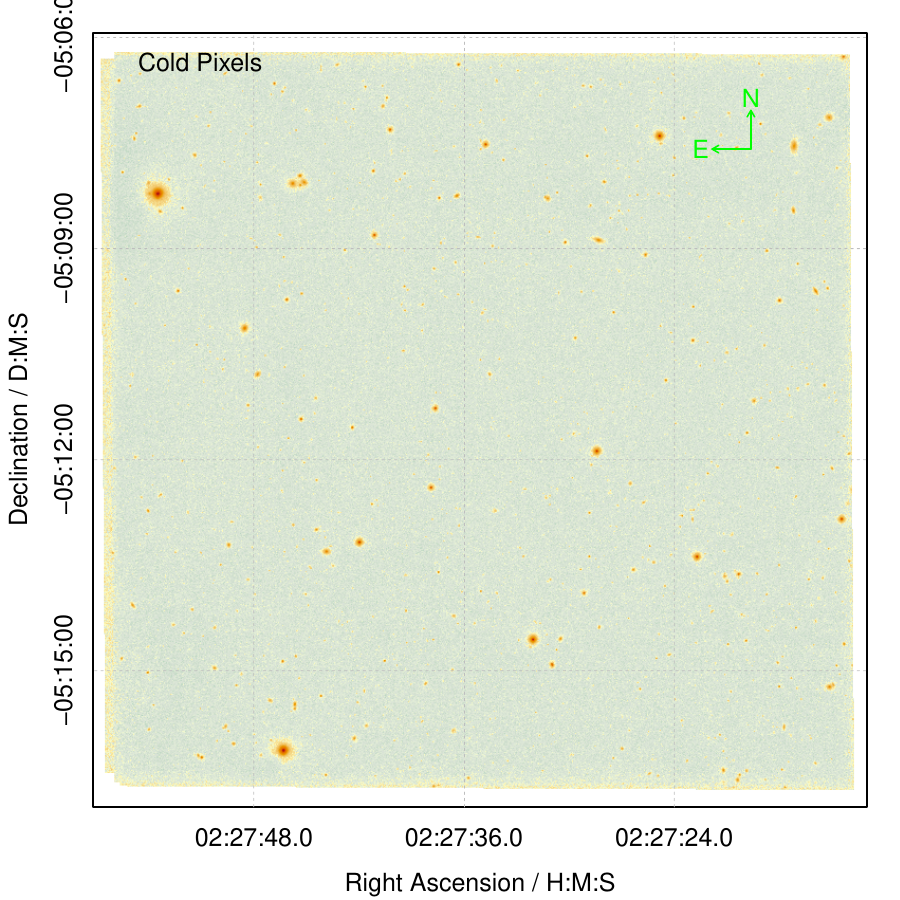}
 \includegraphics[width=\columnwidth]{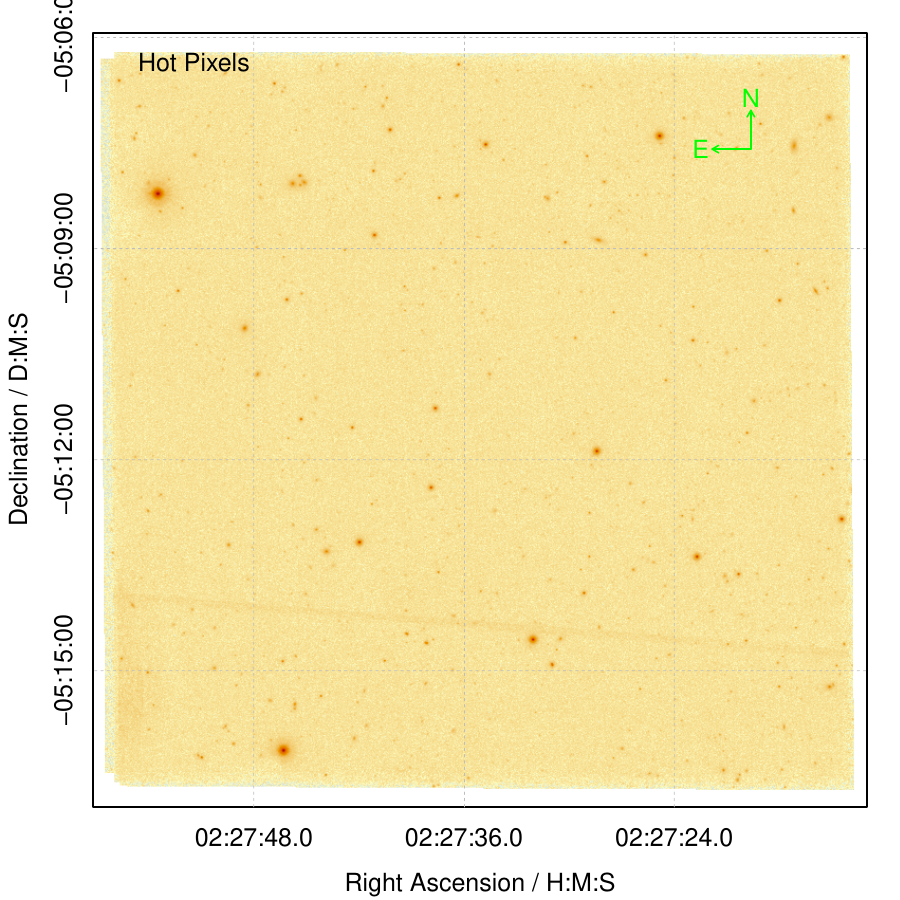} \\
 \caption{Top-left: final inverse variance weighted stacked image using all eight VISTA VIKING frames shown in Figure \ref{fig:input_frames} where the colour scheme is as in Figure \ref{fig:input_frames}. Top-right: the weight map of the stack, where whiter pixels have more contributing frame (up to eight in this case) and blacker pixels fewer contributing frames (as few as zero). Bottom-left: cold pixel values, ie.\ the lowest value across all aligned pixels where the colour scheme is as in Figure \ref{fig:input_frames}. Bottom-right: hot pixel values, ie.\ the highest value across all aligned pixels where the colour scheme is as in Figure \ref{fig:input_frames}. Note where there are no sources the cold pixel image is generally blue (since due to noise at least one pixel is usually negatively valued in the background). For similar but opposite reasons the hot pixel image is generally red everywhere.}
 \label{fig:stack_outputs}
\end{figure*}

\propane{} monitors multiple characteristics of the stacked image. This minimally includes the {\bf image} itself; the {\bf weight} map (how many pixels contribute to the final combined pixel, which ignores masked pixels as well as pixels outside the WCS) and the {\bf inVar} inverse variance weight of the stack. Optionally we can also compute the exposure time map ({\bf exp} how long each pixel was exposed to photons for); a {\bf cold} / {\bf hot} pixel map of lowest and highest values; and a {\bf clip} map (highlighting pixels that appear to be significant outliers after stacking). Figure \ref{fig:stack_outputs} shows the typical inverse variance weighted {\bf image}, {\bf weight} map and {\bf cold} / {\bf hot} pixel images. It is notable that the satellite trail across the bottom of the {\bf hot} pixel image (and present in the bottom row third from left input image shown in Figure \ref{fig:input_frames}) is much fainter in the inverse variance weighted {\bf image}.

\begin{figure}
 \includegraphics[width=1\columnwidth]{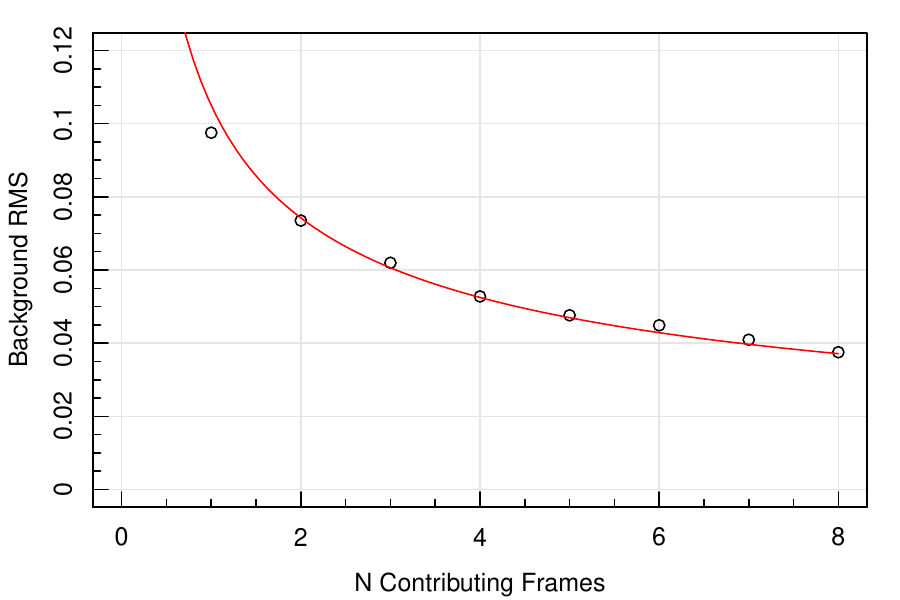}
 \caption{The scaling of the background RMS with the number of contributing VISTA VIKING frames for the stack presented in Figure \ref{fig:stack_outputs}. The curve over-plotted shows the expected square-root scaling as a function of the number of contributing frames.}
 \label{fig:RMS_Nstack}
\end{figure}

We expect the background root mean square (RMS) to reduce with the square-root of the number of contributing frames ($N$) in the case of all input frames having equal depth. When input frames have variable depth this scaling is more generally with the summed inverse variance of the contributing frames to a given pixel. Figure \ref{fig:RMS_Nstack} demonstrates that \propane{} produces the expected scaling. With very large stacks of data it is not unusual for the improvement to scale a bit worse than square-root $N$ because you tend to become limited by image quality systematics rather than pure Normal background noise.

The main alternative mode of stacking (as mentioned above) is median based stacking, where all aligned pixels are assessed and the median value computed. The complexity of this is that there is no exact method to compute medians without all possible values being kept in memory. Approximate methods (using e.g. dynamic histograms) do allow for on-the-fly data incorporation, but generally these are inappropriate for astronomical data since too much needs to be known about the data a-priori to construct the appropriate histogram. The reason this aspect of median stacking becomes an issue is when trying to combine thousands of images simultaneously, i.e.\ it becomes unfeasible to keep perhaps thousands of images projected to a 20k x 20k pixel WCS in memory (this would require many TBs).

The simplest solution for computing pixel medians (and other quantiles) would be to store all projected images in memory, but this might mean hundreds of GBs (or TBs) of memory. This solution obviously becomes unscalable in general. The better route is similar to that utilised in \swarp, where the warped and cropped sub-regions are stored on disk and then small cutout regions are loaded into memory (say $1000 \times 1000$ pixel blocks of a much larger image) and the median statistics computed for all blocks. These blocks can then be reassembled to produce a final median (or other quantile) image. To facilitate this process, \propane{} provides a propaneWarpDump function that warps all input images to a common target WCS and crops them as tightly as possible, saving the outputs to disk with the sub region covered by each cropped image specified in the \FITS{} headers. With these warped and cropped images now on disk, the propaneStackWarpMed function can now loop around the larger imager in smaller blocks ($1000 \times 1000$ pixel blocks being the default) and construct the final median combined image incrementally and in parallel using multiple cores.

\begin{figure*}
 \includegraphics[width=\columnwidth]{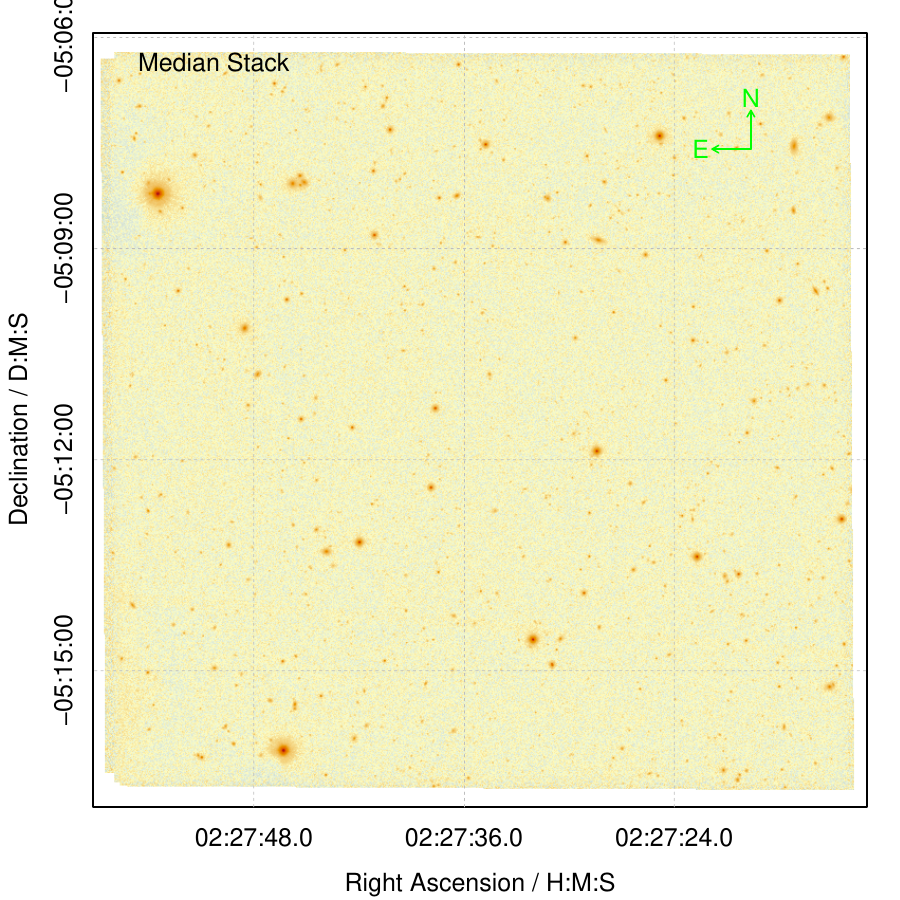}
 \includegraphics[width=\columnwidth]{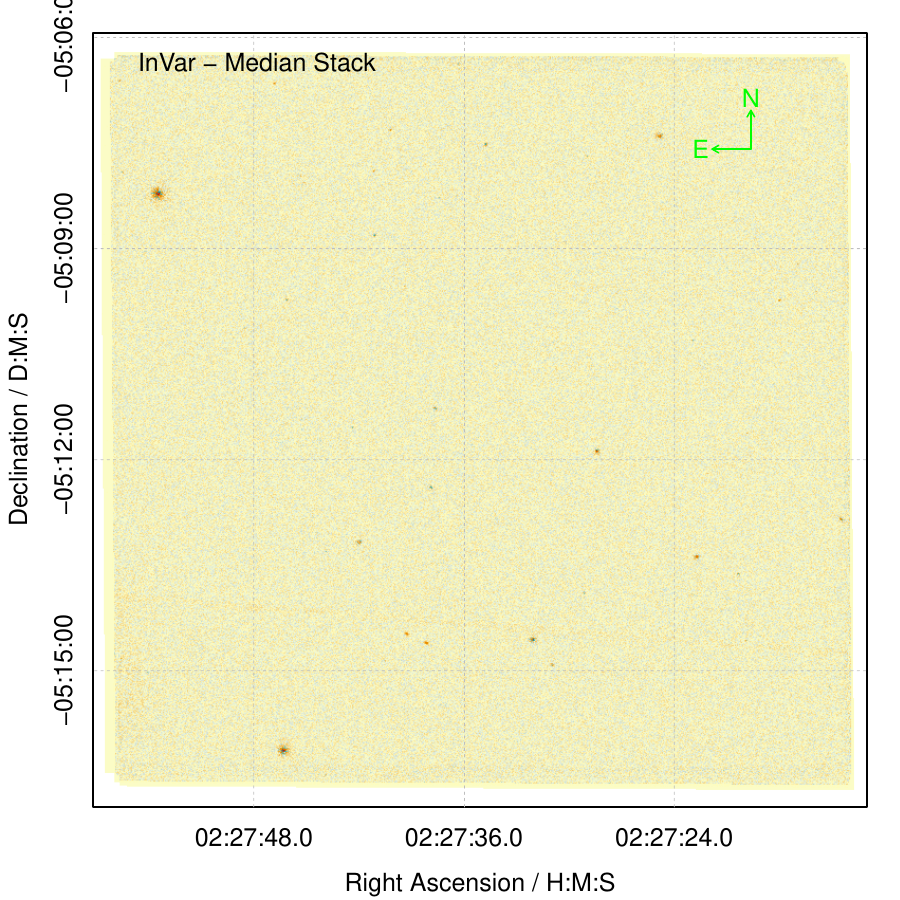}
 \caption{Left: the median combined image. This uses all input images shown in Figure \ref{fig:input_frames}, and can be directly compared to the inverse variance weighted stack presented in Figure \ref{fig:stack_outputs}. Right: the inverse variance weighted stack minus the median combined stack.}
 \label{fig:stack_median}
\end{figure*}

Figure \ref{fig:stack_median} presents the median combined image (LHS) and the inverse variance minus the median combined image (RHS). It is notable that most of the residuals are consistent with noise, except near the cores of very bright objects where pixel saturation creates some larger differences. The satellite trail can be faintly seen as a positive residual towards the bottom of the frame, suggesting it is still present in the inverse variance stacked image.

In general median combined images will have larger pixel RMS than an inverse variance weighted stack. For equal depth images with Normal background statistics the expectation is the median combined RMS should be a factor $\sqrt{\pi/2} = 1.25$ times larger (i.e.\ it is not as deep). In the case above we do find the inverse variance stack is indeed deeper, but the difference in RMS is only a factor 1.13. This might suggest there are some outlier pixels in the inverse variance weighted image preventing the full gain we might expect with truly Normal image statistics, e.g.\ the satellite trail highlighted above.

Since the median combined image approach is very similar to how \swarp{} operates, we can directly compare the stacked images using exactly the same WCS projection scheme. \swarp{} does not offer a bicubic interpolation scheme for pixel remapping, so we used the most similar LANCZOS3 option \citep[see][for discussion on Lanczos interpolation]{citeCoreAlgo}. Depending on the geometry and smoothness of sources, there are potential positives and negatives to either approach. In general Lanczos interpolation techniques work better when images have strong discontinuities and contrast than bicubic approaches.

\begin{figure}
 \includegraphics[width=\columnwidth]{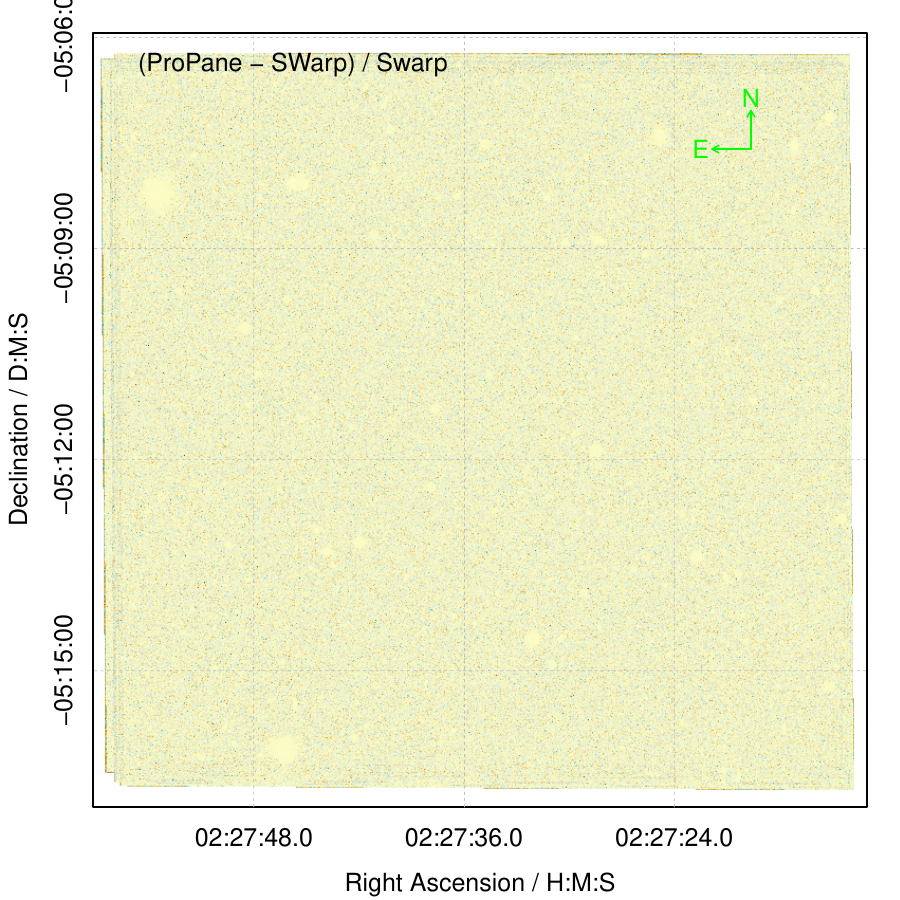}
 \includegraphics[width=\columnwidth]{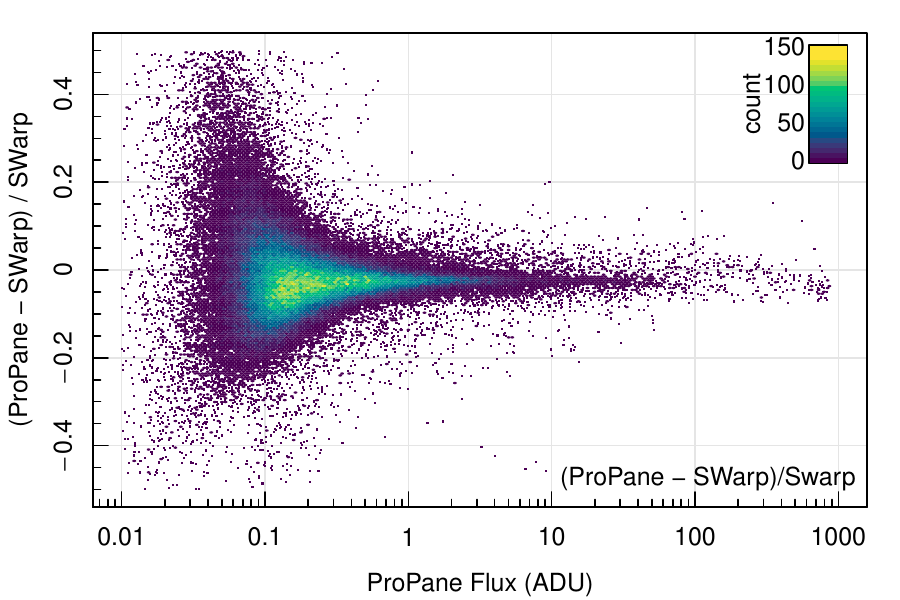}
 \caption{Comparison of \propane{} and \swarp{} median combined stacks (using the same input data as Figure \ref{fig:input_frames}). The top panel shows the relative difference between images, where yellow is very small differences (generally what we see in the location of true sources). The bottom panels shows the per pixel flux differences as a function of \propane{} pixel flux. Again the differences become smaller for brighter (true source) pixels, with a systematic difference of about 2\% (\propane{} pixels contain a bit less flux).}
 \label{fig:propane_swarp}
\end{figure}

The results look most similar where sources clearly exist (since flux varies smoothly in these pixels), and less similar in the background noise (where adjoining pixels can vary a lot in relative flux creating small local discontinuities). This is clear in the top panel of Figure \ref{fig:propane_swarp} comparing directly the \propane{} and \swarp{} median combined stacks. The bottom panel of Figure \ref{fig:propane_swarp} does show some small systematic differences, where \propane{} tends to have about 2\% less flux than the matching \swarp{} pixels even for high $S/N$ pixels. This difference is likely due to the known behavioural differences between Lanczos and bicubic interpolation, and is generally similar to or less than typical zero-point errors, and much less than photometric measurement systematics \citep[e.g.\ see figure 14 of][]{2020MNRAS.496.3235B}.

Some of the brightest pixels in Figure \ref{fig:propane_swarp} show particularly discrepant behaviour because the pixels are saturated and produce locally sharp features. Such sharp features exacerbate the algorithmic differences between \swarp{} and \propane. In general, the scatter and offset is similar to that found in the warping tests conducted at the end of Section \ref{sec:image_warping}. If photometry more accurate than a couple of percent in flux is desired, then in general users should not be stacking / combing data at all. Instead one should use native projection images and the segmentation map projection method discussed above.

\subsection{Bad Pixels and Image Artefacts}

An important component of image combining is flagging and removing (or patching) `bad' pixels. Generally a bad pixel will be one where the flux measurement is compromised, this might be due to cosmic rays hitting the detector, satellite trails in the image, or hardware issues in the detector itself (dead pixels etc). Many facilities have mechanisms to flag bad pixels and include this information in an extension of the \FITS{} image (HST and JWST have particularly comprehensive pixel flagging), but it is also common that specific image issues are not flagged at all and the end user must identify and deal with these artefacts. \propane{} includes a number of mechanisms to deal with such image artefacts, and in general these will need to be used before trying to combine images together to make a stacked image. It should be noted that some image stacking techniques, particularly median combining, are more resilient to the presence of image artefacts, but even in this situation further improvements will come from proper flagging of bad pixels.

\subsubsection{Outlier Pixel Flagging}

The first method of pixel flagging is an iterative approach during the image stacking, where the cold and hot (lowest and highest valued contributing pixels to a given stacked pixels) are monitored. After the initial stack of the image, cold and / or hot pixels that are in tension with the final stacked value (controlled with user-defined parameters) are flagged as being `bad', and a second round of stacking is done where these flagged pixels are entirely masked. Cold and hot pixels from a single round of stacking (for eight input VISTA VIKING images) can be seen in the bottom two panels of Figure \ref{fig:stack_outputs}. Note how the cold pixels will usually sit below the background (appearing blue) and hot pixels will generally sit above (appearing red). In this example we can clearly see a satellite trail as an image artefact in the hot pixels, but not present in the cold pixels (where other bright sources are still present).

Outlier pixel detection works well for flagging bad pixels, but it does have some disadvantages. Firstly, it roughly doubles the time taken to stack the image, since the second round of stacking generally requires a complete re-warping of all input images. It is possible to save the intermediate warped images on disk (significantly speeding up this second round of stacking), but that obviously increases the amount of disk IO substantially. Secondly, such an iterative clipping method works best when there are a lot of contributing data in a given pixel. This is obvious at the extreme, because if there are only two images contributing to a given pixel, then whilst they might have significantly different values for that pixel we do not know which one is the actual outlier --- i.e.\ is our final pixel value too low / cold or too high / hot? Generally, this approach to flagging bad pixels works well when we have three or more frames contributing to a pixel. It is also possible that multiple images provide overlapping `hot' pixels, e.g.\ two satellite trails from two different images that happen to overlap in the sky. To combat this efficiently we use image patching (discussed below).

\subsubsection{Bad Pixel Flagging}

To account for the limitations of iterative stack-based bad pixel masking, \propane{} also includes the propaneBadPix routine that works on individual images. Conceptually, this function operates by creating an unsharp mask version of the image (subtracting a slightly smoothed version of an image from itself), and then comparing the unsharp mask to the RMS levels of the image. The reason this works is real astrophysical images tend to not have very sharp spatial discontinuities, with bright fluxes generally changing smoothly over at least a few pixels (for a well-sampled image). If a strong discontinuity is present in the unsharp mask (relative to the expected RMS of the image) then these pixels are flagged as being bad image artefacts.

To improve the quality of later stacking, these flagged bad pixel regions can be further dilated (grown) to capture additional contiguous pixels. This is often desirable behaviour, because the edges of the bad pixel regions can be lost when constructing the unsharp mask. For additional flexibility, this function can operate to only detect cold and / or hot pixels. Some detector types suffer from cold and hot pixels equally (e.g.\ near infrared) whilst other might only experience significant hot pixel artefacts (e.g.\ cosmic rays in space telescopes in particular). Even an event that clearly causes strongly hot artefacts (like a cosmic ray) can produce adjacent regions of cold pixels due to electronic errors and read out processing, so users should be careful to inspect their images to ensure the correct combination of bad pixel flagging is used. There are a number of parameters that control how propanePatchPix operates, so interested users should refer to the extensive documentation and included examples.

\subsubsection{Image Patching}

\propane{} offers two methods to further patch images to improve image quality: pixel patching using a smoothed image (used on a single frame), and combining inverse variance weighted and median combined stacks optimally.

First we will discuss pixel patching via the propanePatchPix function. Even in situations where bad pixels have been correctly flagged, users might not simply want to mask out these regions entirely when combining images. This is particularly the case when not many frames exist in a given region of sky, and we want accurate photometry for any measured sources. If we use very aggressive bad pixel flagging and masking, then these regions often have missing pixels even after stacking. In this case, small sources in particular might have biased fluxes without some attempt to patch in missing flux values. Also, colour images created in these regions can have quite distracting colour artefacts (since pixels might only be missing in some colour channels).

In some situations, especially when bad pixels exist inside the smooth flux profiles of real sources, it is perfectly possible to infer reasonable values for the bad pixel based on flux values of local pixels. \propane{} offers a basic method to patch in the value of bad pixels, where the missing values are reconstructed from a smoothed version of the image with the bad pixels masked out. The smoothing kernel can be modified, but generally the image should be smoothed on a scale of 1--3 pixels (so only very local values are used to infer the missing pixel flux).

The second method of patching (propanePatch) requires the user to have both inverse variance weighted and median combined stacks. A number of options allow the user to determine what degree of difference between the two is considered concerning (based on the flux offset given the predicted variance of the image at each pixel). Where the two images appear to be in tension, the solution for the median combined image can be patched in. This method works for both relatively `cold' and `hot' pixels in the inverse variance weighted stack. In general only a fraction of a percent of pixels are flagged for median patching even with quite aggressive settings, and these pixels usually correspond to small amount of artefacts (like the satellite trails discussed above) and the saturated cores of bright stars (which have ill-defined flux regardless).

\section{Applications of \propane}
\label{sec:applications}

Below we briefly discuss some recent literature examples of \propane{} being used to produce scientific ready imagine products.

\subsection{WAVES Deep Drill Fields}

The Wide Area VISTA Extragalactic Survey \citep[WAVES;][]{2019Msngr.175...46D} is an upcoming large spectroscopic programme operating on the 4MOST facility \citet{2019Msngr.175....3D}. The main original survey comprises of three different sub surveys: WAVES Wide (covering much of the VST KiDS survey area, see Table \ref{tab:acronym} for survey description); WAVES Deep G23 (largely targeting the GAMA G23 field, see Table \ref{tab:acronym} for survey description); and WAVES Deep Drill Fields (DDFs, targeting four well known deep fields that will also be targeted by LSST\footnote{Large Survey of Space and Time} with the Vera C. Rubin Observatory).

To create the input catalogues for the WAVES DDFs a large number of images from different surveys had to be combined to ensure we covered the full area of the 4MOST field-of-view. To allow the selection of sources using photometric redshifts images had to be created and analysed covering $ugri$ VST optical (KiDS; see Table \ref{tab:acronym} for survey description) and ZYJHKs VISTA near-infrared (NIR) bands (VIKING; see Table \ref{tab:acronym} for survey description). In the case of the NIR data, additional fields had to be observed with VISTA to guarantee we had full area coverage (this data was collected between 2021--2023). The large variation of individual frame depth (in terms of exposure time) and quantity of overlapping data necessitated the construction of inverse variance weighted stacks, and we used \propane{} to create our deep square-degree stacks (a square degree being a reasonable upper limit for the resulting \FITS{} file size). Across all DDF regions and bands, 522 individual square-degree stacks had to be constructed for later photometric analysis.

\begin{figure}
 \includegraphics[width=\columnwidth]{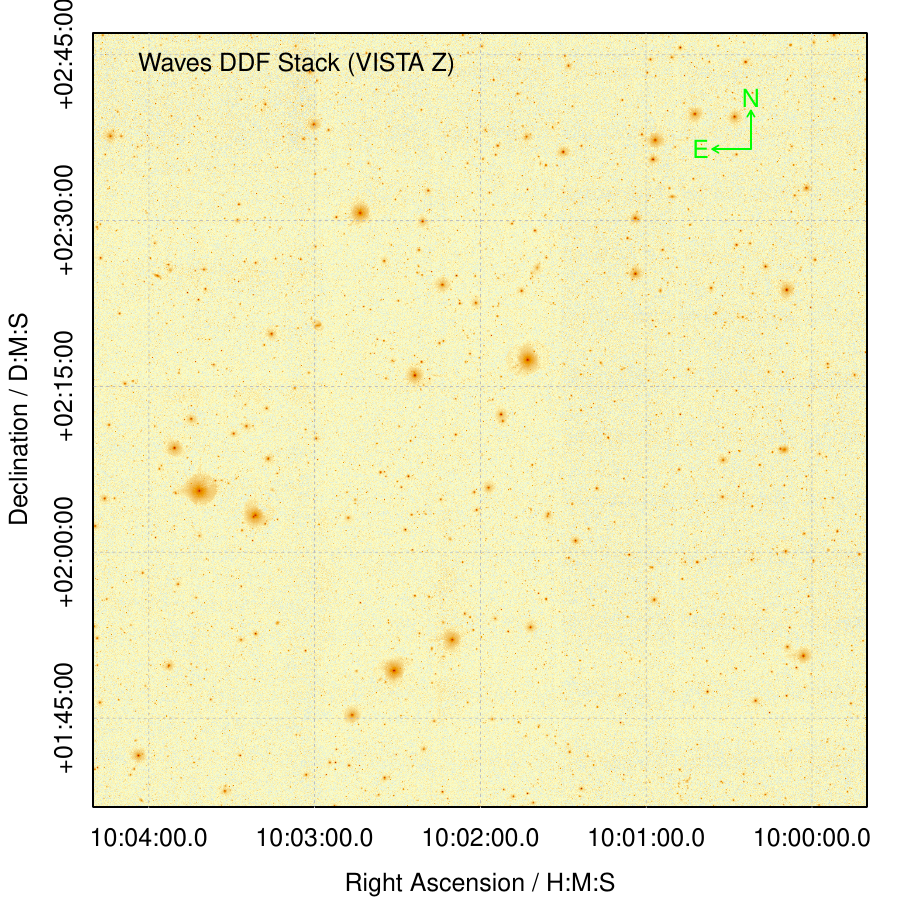}
 \includegraphics[width=\columnwidth]{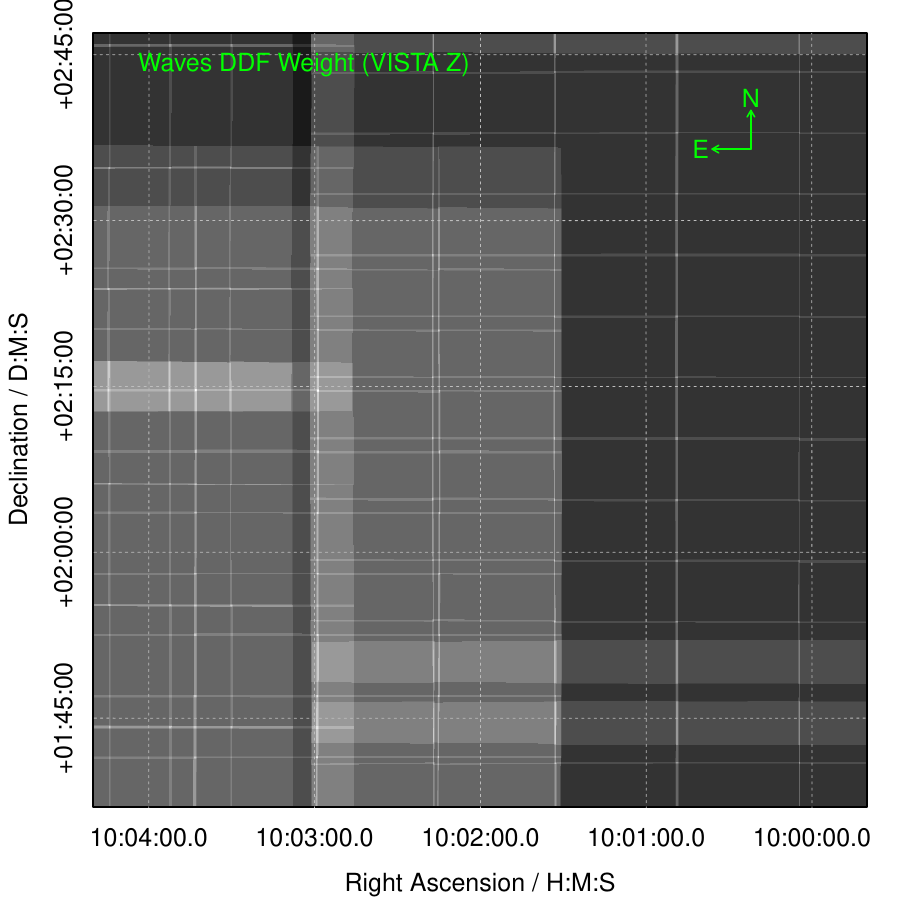}
 \caption{Top: inverse variance weighted Z-band stack of a square degree region in the WAVES Deep Drill Field WD10, where the colour scheme is as in Figure \ref{fig:input_frames}. Bottom: weight map of the 164 contributing images that created the stack, where lighter colours represents more frames contributing to a pixel.}
 \label{fig:propane_DDF}
\end{figure}

Figure \ref{fig:propane_DDF} shows an example square degree in the VISTA Z-band. The data varies a lot in depth across the area shown (as seen in the top panel of the Figure), with the modal number of frames contributing to a pixel being two, the median being three, and the maximum being 10 (although only 1701 out of 196 million pixels have that degree of overlap). Because \propane{} correctly monitors the inverse variance depth of the image at every pixel, the subsequent source extractions using \profound{} is robust to very large changes in the apparent stack depth. This is a particular issue in fields that overlap with the very deep VIDEO and UltraVISTA surveys (see Table \ref{tab:acronym} for survey descriptions), where we can have sharp transitions from singular frames (from our additional VISTA coverage programme) to hundreds of frames contributing to stacked pixels.

\subsection{James Webb Space Telescope}

Some of the features of \propane, specifically its flexibility regarding distortion terms, were developed with JWST data in mind. It has been used extensively as part of the large Guaranteed Time Observation programme PEARLS \citep[see][]{2023AJ....165...13W, 2023arXiv231003081D}, where it has been a useful tool for stacking images and aligning them with their HST counterparts (e.g.\ see Figure \ref{fig:tweak_ex}).

\begin{figure}
 \includegraphics[width=\columnwidth]{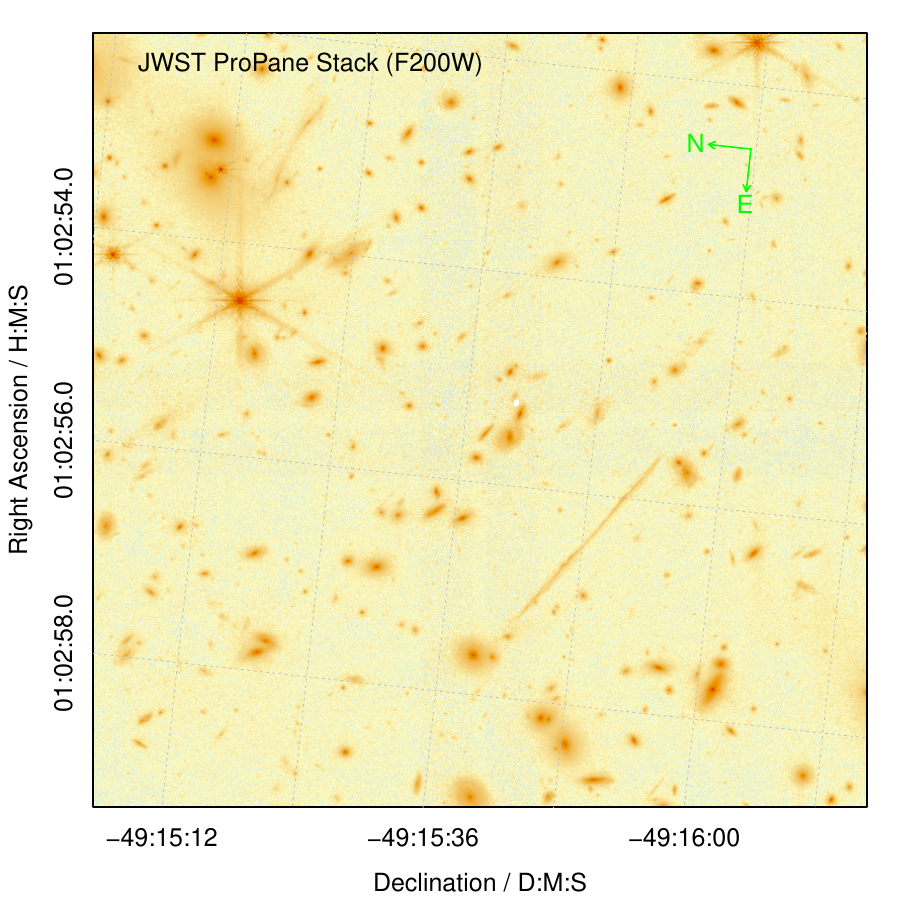}
 \includegraphics[width=\columnwidth]{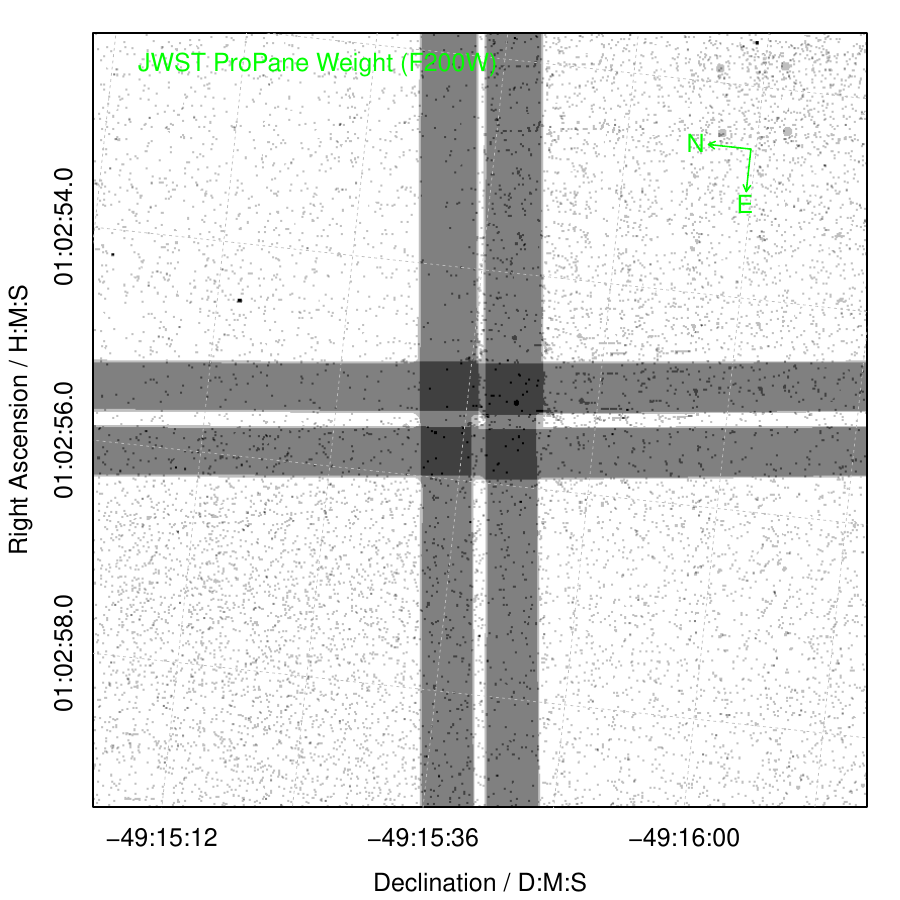}
 \caption{Top: inverse variance weighted JWST NIRCam F150W stack of the El Gordo Cluster Field, where the colour scheme is as in Figure \ref{fig:input_frames}. Bottom: weight map of the 16 contributing images that created the stack, where lighter colours represents more frames contributing to a pixel.}
 \label{fig:propane_JWST}
\end{figure}

Figure \ref{fig:propane_JWST} is an example \propane{} inverse variance stack using the F200W filter in the El Gordo Cluster Field that was targeted as part of the PEARLS medium-deep programme \citep{2023AJ....165...13W}. Particularly notable compared to the stacked image shown in Figure \ref{fig:propane_DDF} is how many pixels tend to be masked during stacking (small patches of variable depth throughout the weight map). In general this is because we use conservative pixel masking options from the JWST DQ extension that quite aggressively flags cosmic rays and other image artefacts that are particularly prevalent in JWST NIRCam images. This is partly why so much dithering is key to making effective use of JWST (otherwise there would be many occurrences of entirely missing data).

\begin{figure}
\includegraphics[width=\columnwidth]{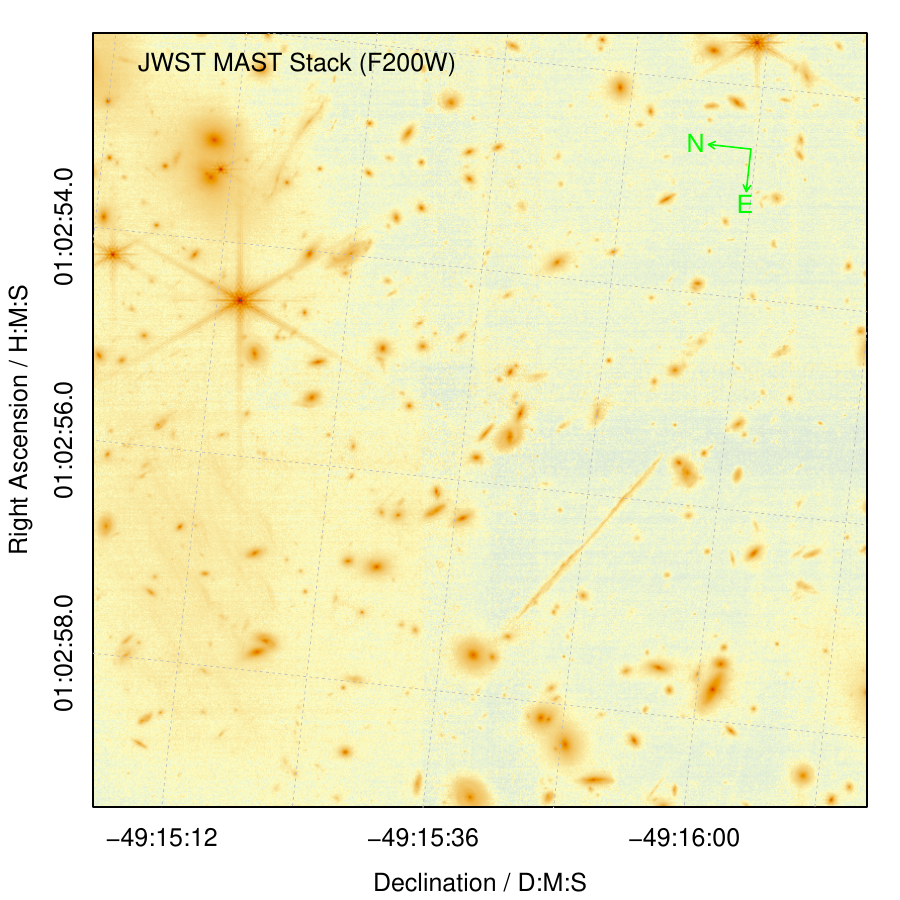}
 \includegraphics[width=\columnwidth]{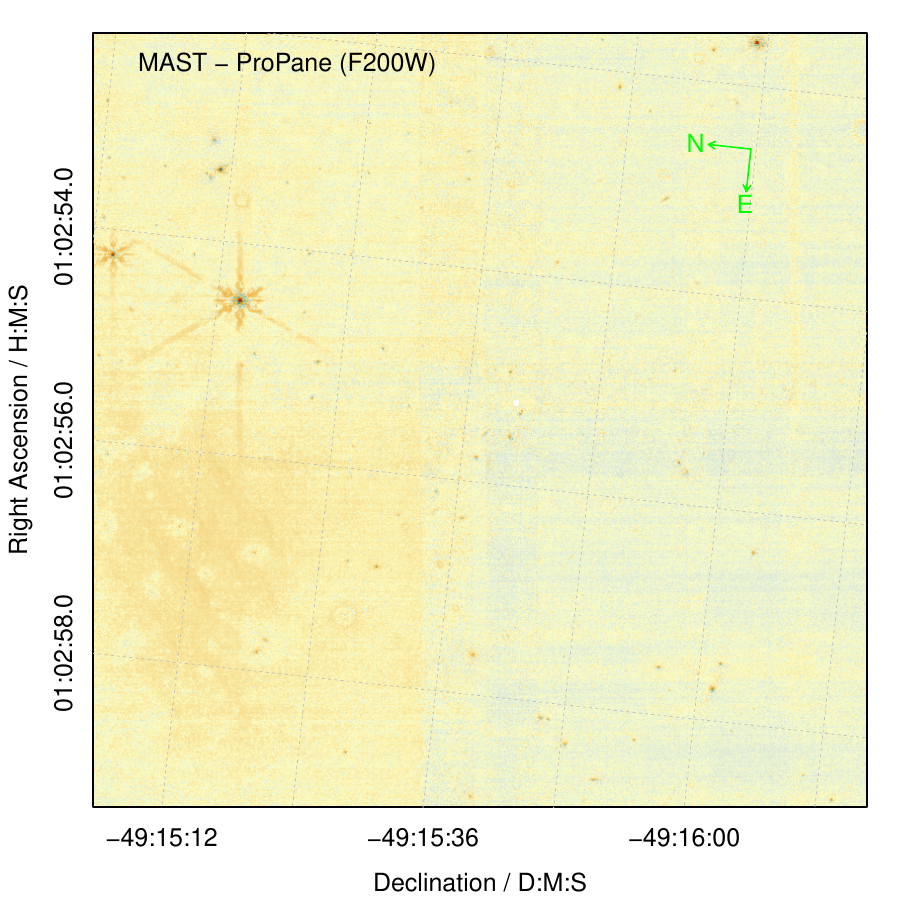}
 \caption{Top: The JWST MAST pipeline El Gordo drizzled image showing the same region as Figure \ref{fig:propane_JWST}. Bottom: MAST - \propane{} of the same region. It is notable that discontinuities, wisps and $1/f$ noise are more apparent in the MAST pipeline version.}
 \label{fig:MAST_JWST}
\end{figure}

It is instructive to compare the \propane-constructed F200W JWST image with the drizzled mosaic available directly from MAST (created with AstroDrizzle). The same region as shown in Figure \ref{fig:propane_JWST} is extracted from MAST and presented in Figure \ref{fig:MAST_JWST}. It is notable that the MAST pipeline version of this region shows much stronger discontinuities from the overlapping frames near the centre, i.e.\ features that line up with the weight map edges presented in the bottom panel of Figure \ref{fig:propane_JWST}. We also see strong $1/f$ noise (short horizontal features) and a prominent wisp in the bottom-left of the frame. The lack of wisps in the \propane{} version of the stack is directly due to the wisp mitigation strategy outlined in \citet{2023PASP..135h5003R} and applied to PEARLS data. The reduced discontinuities and $1/f$ reduction in the \propane{} stack are due to the \profound{} pipeline used in PEARLS, as outlined in \citet{2023AJ....165...13W}.

\begin{figure}
 \includegraphics[width=\columnwidth]{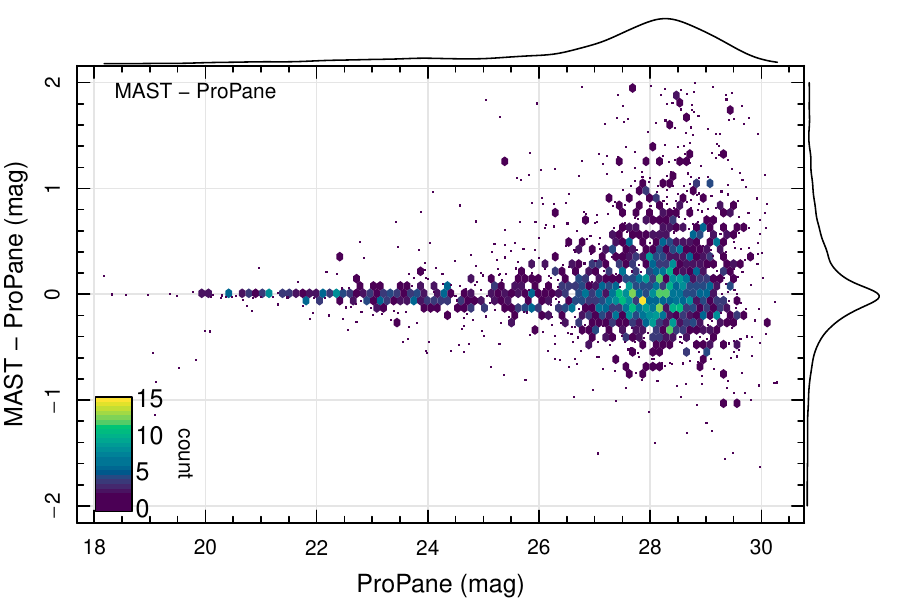}
 \caption{Comparison of course magnitudes for the same segments, applied to \propane{} and MAST images shown in Figures \ref{fig:propane_JWST} and \ref{fig:MAST_JWST}.}
 \label{fig:MAST_ProPane_mag}
\end{figure}

The actual source magnitudes are directly compared in Figure \ref{fig:MAST_ProPane_mag}. \profound{} was initially run on the \propane{} version of the image (i.e.\ top panel of Figure \ref{fig:propane_JWST}) and the same segments were then applied in forced photometry mode to the MAST AstroDrizzle mosaic. In both cases \profound{} was allowed to determine the optimum sky subtraction using a box filter that was one quarter the dimension of the image. Figure \ref{fig:MAST_ProPane_mag} shows fair agreement in the measured flux for brighter sources, particularly those above $26^{th}$ mag. Beyond this we find a large degree of random scatter, where most of this is due to sky subtraction differences locally. This is not surprising, given the clear difference in sky flatness noted between the \propane{} stack and the MAST Drizzle (with the PEARLS \propane{} pipeline producing a much flatter and artefact-free product). In general, the flux similarity of the \profound{} source extraction gives confidence that the astrometric projection and stacking is largely similar between the MAST drizzle pipeline and \propane

\subsection{Hyper Suprime Camera}

\propane{} can also be used to combine images without any associated warping. Within \propane{} this is referred to as `flat' stacking (as opposed to `warp' stacking in which a WCS is specifically targeted). Where this might be useful is when stacking targeted sources with `similar' properties to create a more reliable measurement. This procedure was used with Hyper Suprime Camera (HSC) data to create large angular size images of the point spread function (PSF). Individual images of the PSF will not contain all the information we need: bright stars are saturated in the cores, and faint stars contain very little flux at large distances from the centre. By combining a mixture of star brightnesses and scaling the images appropriately, a very deep image can be constructed which contains fine resolution detail in the centre, and low surface brightness information at large radii.

\begin{figure}
 \includegraphics[width=\columnwidth]{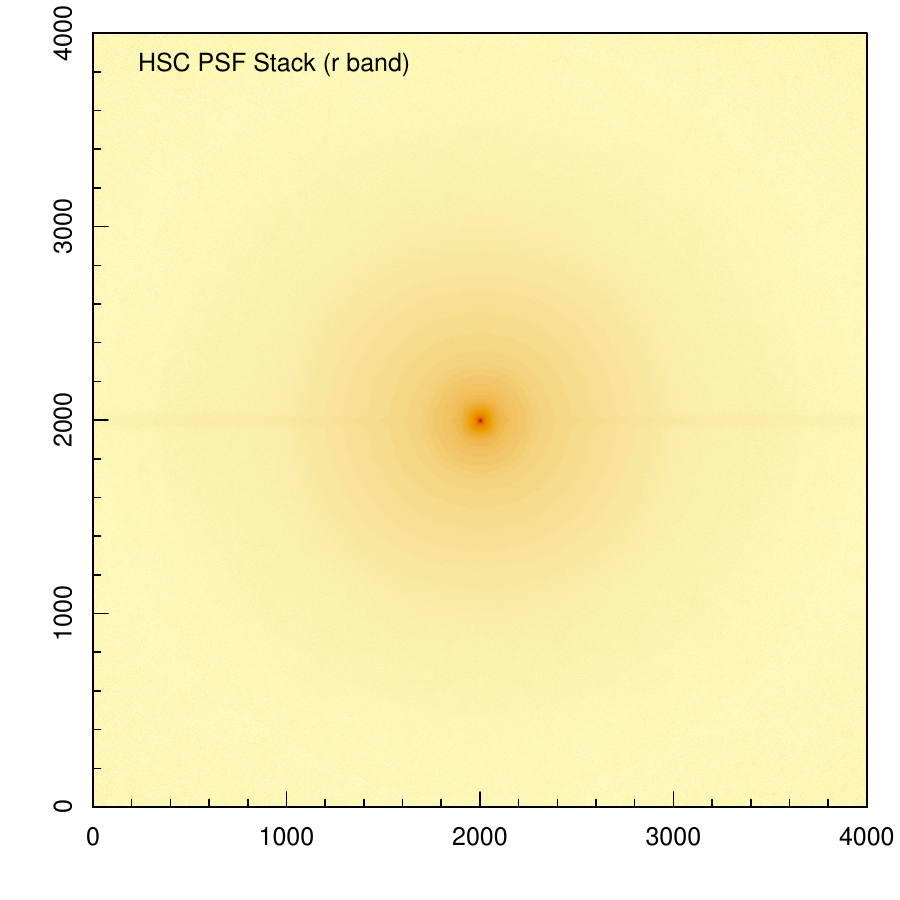}
 \includegraphics[width=\columnwidth]{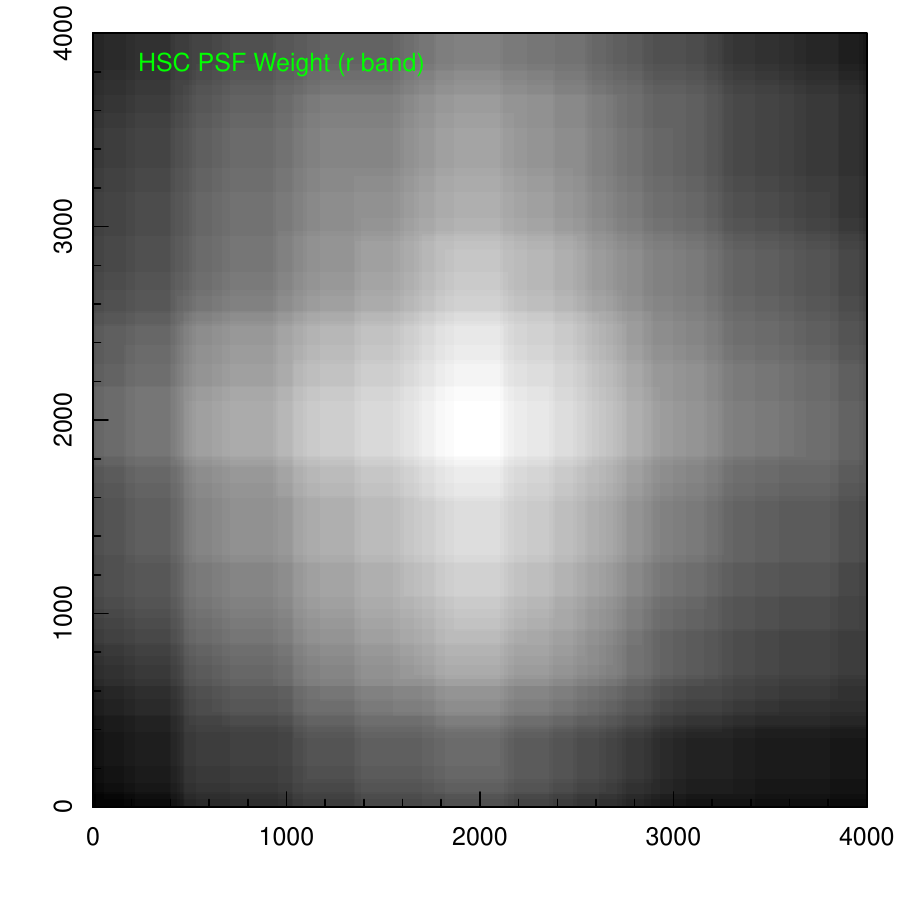}
 \caption{Top: inverse variance weighted HSC $r$-band stack of PSFs, where the colour scheme is as in Figure \ref{fig:input_frames}. Bottom: weight map of the 86 contributing bright star images that created the stack, where lighter colours represents more frames contributing to a pixel.}
 \label{fig:propane_HSC}
\end{figure}

This procedure is presented in detail in \citet{2023arXiv230916244G}, but to summarise the usage of \propane{} Figure \ref{fig:propane_HSC} shows the final $r$-band HSC PSF constructed, with the variable depths of the contributing images presented in the bottom panel. The data was median combined to make the final image of the PSF, with no requirement to even mask other sources since surrounding sources of random stars and galaxies will not be consistently aligned and will disappear when computing the per pixel median. Given the targeting of known stars the image is by construction deepest in the centre, with fewer frames contributing on average as we move further out. Combining stars like this is extremely fast, since the image warping typically dominates the computation time (not the final combining/stacking process).

\section{Conclusions}
\label{sec:conclusions}

In this work we formally introduce \propane{} to the literature and briefly discuss the new associated packages \Rfits{} and \Rwcs{} which offer interfaces to the \CFITSIO{} and \wcslib{} \C{} libraries for \FITS{} IO and WCS manipulation respectively.

\propane{} is a new image warping and combining software package written in \R{} that is particularly built on the \Rwcs{} and \imager{} packages (and by association the \wcslib{} and \CImg{} \Cpp{} libraries they use) that allows for astronomy specific image IO, projections and manipulation. \propane{} has access to a large variety of projection and distortion terms (all those included in \wcslib) and offers numerous strategies to warp (e.g.\ forwards/backwards; bilinear/bicubic) and can scale to stacking an arbitrary number of images. It is a naturally multicore code, and includes threading for many of its tasks via the \imager{} library.

As well as the core task of image warping and stacking, \propane{} offers a number of important utility routines that we summarise below:

\begin{itemize}
\item {\bf Combining} --- numerous methods to combine images, largely via the \imager{} package. This includes inverse variance and median combining, as well as the extraction of higher order statistics from image stacks.
\item {\bf Finding} --- efficient routines to determine which images potentially overlap with a given region of sky or WCS.
\item {\bf Masking} --- image and stack based routines to mask potentially bad pixels, allowing us to remove them from subsequent combining and stacking processes.
\item {\bf Tweaking} --- catalogue and pixel-based routines to align images with slightly mismatching WCS. This allows for both translation and rotation related tweaks (simultaneously).
\item {\bf Warping} --- general purpose warping between WCS (and manual warping also). All modern projection and distortion terms are accommodated via the \wcslib{} library and related \Rwcs{} package.
\end{itemize}

It has been used in a number of previous works (see Section \ref{sec:applications}), where it has successfully produced very deep image stacks for current (e.g.\ PEARLS) and future (e.g.\ WAVES) surveys. \propane{} will be maintained and developed in an ongoing manner, as has been the case for other software constructed as part of the \protools{} software suite.

The main future development direction for \propane{} will be the addition of full PSF homogenisation. Usually this means trying to convolve images with a suitable kernel that means all images have the same effective PSF. In practice the easiest route is to convolve images to the `worst' (broadest) PSF since de-convolving an image (sharpening it, in effect) is a computer science problem fraught with difficulties. Even small errors in estimating the true image PSF can create significant image artefacts. Adding to this the reality that PSFs vary across a frame due to telescope optics, and it is clear much care must be taken.

\section*{Acknowledgements}

ASGR acknowledges funding by the Australian Research Council (ARC) Future Fellowship scheme (FT200100375). SPD acknowledges funding by the Australian Research Council (ARC) Laureate Fellowship scheme (FL220100191). LJD acknowledges funding by the Australian Research Council (ARC) Future Fellowship scheme (FT200100055).


\section*{Data Availability}

\propane{} v1.6 is released for this paper on GitHub using a permissive LGPL-3 license \citep[github.com/asgr/ProPane;][]{ProPane}. All associated \protools{} software (v0.2) is available on GitHub for use immediately \citep[github.com/asgr/ProTools;][]{ProTools}.



\bibliographystyle{mnras}
\bibliography{ProPane} 



\appendix

\section{Glossary}

For easy reference, common acronyms (including surveys and telescopes) and software are expanded and described in Tables \ref{tab:acronym} and \ref{tab:software}. Note that until 2020 `LSST' was the name of the facility, i.e.\ the Large Synoptic Survey Telescope. This facility is now known as the Vera C. Rubin Observatory, and LSST has been re-purposed as the name of the main survey to be conducted (as detailed in Table \ref{tab:acronym}).

\begin{table*}
\caption{Quick reference for common astronomy acronyms, survey and telescopes used throughout this paper.}
\begin{center}
\begin{tabular}{|l|l|l|l|}

Acronym & Expanded & Explanation & Reference (if relevant) \\

\hline

4MOST & 4m Multi-Object Spectroscopic Telescope & 2400 multi object spectrograph upgrade to VISTA & \citet{2019Msngr.175....3D} \\
COSMOS & Cosmic Evolution Survey & Multi-facility survey targeting a $\sim$2 square degree field & \citet{2007ApJS..172..196K} \\
CPU & Central Processing Unit & Processing core on a computer (usually one of many) & NA \\
FWHM & Full width half maximum & PSF size: the full measured width at half peak brightness & \citet{2017MNRAS.466.1513R} \\
GALEX & Galaxy Evolutionary Explorer & UV space telescope & \citet{2007ApJS..173..682M} \\
GAMA & Galaxy And Mass Assembly & Major spectroscopic survey on the Anglo Australian Telescope & \citet{2011MNRAS.413..971D} \\
HST & Hubble Space Telescope & 2.4m optical to NIR space telescope & NA \\
JWST & James Webb Space Telescope & 6.5m NIR to MIR space telescope & NA \\
LSST & Large Survey of Space and Time & Major optical survey at the Vera C. Rubin Observatory & \citet{2019ApJ...873..111I} \\
NIR & Near Infrared & Wavelength range covering approximately 800--2500 nm & NA \\
MIR & Mid Infrared & Wavelength range covering approximately 2.5--25 micron & NA \\
PEARLS & \makecell[l]{Prime Extragalactic Areas for \\ Reionization and Lensing Science} & Major JWST programme & \citet{2023AJ....165...13W} \\
PSF & Point Spread Function & Profile of a point source (often a star) in an image & NA \\
SED & Spectral Energy Distribution & Electromagnetic radiation of an astronomical source & \citet{2020MNRAS.495..905R} \\
UltraVISTA & Ultra Deep Survey with the VISTA Telescope & NIR survey of COSMOS on VISTA & \citet{2012AandA...544A.156M} \\
UV & Ultra Violet & Wavelength range covering approximately 100--400 nm & NA \\
VIDEO & VISTA Deep Extragalactic Observations Survey & Large deep NIR survey on VISTA & \citet{2013MNRAS.428.1281J} \\
VIKING & VISTA Kilo-Degree Infrared Galaxy Survey & Large wide NIR survey on VISTA & \citet{2013Msngr.154...32E} \\
VISTA & \makecell[l]{Visible and Infrared Survey \\ Telescope for Astronomy} & 4m NIR telescope located at Paranal Observatory (Chile) & NA \\
VST & Very Large Telescope Survey Telescope & 2.6m optical telescope located at Paranal Observatory (Chile) & NA \\
WAVES & Wide Area VISTA Extragalactic Survey & Major spectroscopic survey on 4MOST & \citet{2019Msngr.175...46D} \\
WCS & World Coordinate System & Converts image pixels to celestial coordinates & NA \\

\end{tabular}
\end{center}
\label{tab:acronym}
\end{table*}%

\begin{table*}
\caption{Quick reference for common astronomy software used throughout this paper.}
\begin{center}
\begin{tabular}{|l|l|l|l|}

Name & About & Reference & Available \\

\hline

\C & Low level general purpose language & NA & www.iso.org/standard/74528.html \\
\Cpp & Low level object oriented language & NA & isocpp.org \\
\CFITSIO & \C{} library for \FITS{} IO & \citet{1999ASPC..172..487P} & heasarc.gsfc.nasa.gov/fitsio/fitsio.html \\
\CImg & Cool Image, a \Cpp{} library for image manipulation operations & \citet{citeCImg} & cimg.eu \\
\drizzle & Pipeline to mosaic and stack images & \citet{2002PASP..114..144F} & www.stsci.edu/scientific-community/software \\
\FITS & Flexible Image Transport System file format standard & NA & \citet{1999ASPC..172..487P}  \\
\HDF & Hierarchical Data Format 5 file format standard and \C{} library & NA & www.hdfgroup.org/solutions/hdf5/ \\
\imager & \R{} package for image manipulation using \CImg & NA & github.com/asgr/imager \\
\profit & \R{} package for 2D galaxy profiling & \citet{2017MNRAS.466.1513R} & github.com/ICRAR/ProFit \\
\profound & \R{} package for source extraction & \citet{2018MNRAS.476.3137R} & github.com/asgr/ProFound \\
\profuse & \R{} package for spatial SED modelling  & \citet{2022MNRAS.513.2985R} & github.com/asgr/ProFuse \\
\propane & \R{} package for warping and combining images & This paper; \citet{ProPane} & github.com/asgr/ProPane \\
\prospect & \R{} package for global SED modelling & \citet{2020MNRAS.495..905R} & github.com/asgr/ProSpect \\
\protools & Suite of \R{} packages for astronomy & \citet{ProTools} & github.com/asgr/ProTools \\
\R & High-level programming language focussed on data analysis & \citet{citeR} & www.r-project.org \\
\Rfits & \R{} package for \FITS{} file IO using \CFITSIO & NA & github.com/asgr/Rfits \\
\Rwcs & \R{} package for WCS conversion using \wcslib & NA & github.com/asgr/Rwcs \\
\swarp& \C{} software to mosaic and stack images & \citet{2010ascl.soft10068B} & www.astromatic.net/software/swarp/ \\
\wcslib & \C{} library for WCS conversion & \citet{2002AandA...395.1077C} & www.atnf.csiro.au/people/mcalabre/WCS/ \\

\end{tabular}
\end{center}
\label{tab:software}
\end{table*}%

\section{Projection Options}

\begin{table}
\begin{center}
\begin{tabular}{l|l}

RA & 	Right Ascension \\
DEC &	Declination \\
GLON &	Galactic Longitude \\
GLAT &	Galactic Latitude \\
ELON &	Ecliptic Longitude \\
ELAT &	Ecliptic Latitude \\
HLON &	Helioecliptic Longitude \\
HLAT &	Helioecliptic Latitude \\
SLON &	Supergalactic Longitude \\
SLAT &	Supergalactic Latitude

\end{tabular}
\end{center}
\caption{Axis types recognised by \Rwcs, and by extension \propane.}
\label{tab:axis_type}
\end{table}%

\begin{table}
\begin{center}
\begin{tabular}{l|l}

AZP &	Zenithal / Azimuthal Perspective \\
SZP &	Slant Zenithal Perspective \\
TAN &	Tangent Gnomonic \\
STG &	Stereographic \\
SIN &	Orthographic / Synthesis \\
NCP &	Unofficially Supported SIN-like Projection  \\
ARC &	Zenithal / Azimuthal Equidistant \\
ZPN &	Zenithal / Azimuthal Polynomial \\
ZEA &	Zenithal / Azimuthal Equal Area \\
AIR &	Airy Projection \\
CYP &	Cylindrical Perspective \\
CEA &	Cylindrical Equal Area \\
CAR &	Plate Carree \\
MER &	Mercator Projection \\
COP &	Conic Perspective \\
COE &	Conic Equal Area \\
COD &	Conic Equidistant \\
COO &	Conic Orthomorphic \\
SFL &	Sanson-Flamsteed (AKA `Global Sinusoid') \\
PAR &	Parabolic \\
MOL &	Mollweide's Projection \\
AIT &	Hammer-Aitoff \\
BON &	Bonne Projection \\
PCO &	Polyconic \\
TSC &	Tangential Spherical Cube \\
CSC &	COBE Quadrilateralized Spherical Cube \\
QSC &	Quadrilateralized Spherical Cube \\
HPX &	HEALPix \\
XPH &	HEALPix Polar

\end{tabular}
\end{center}
\caption{Projection types supported in \Rwcs, and by extension \propane.}
\label{tab:proj_type}
\end{table}%

\begin{table}
\begin{center}
\begin{tabular}{l|l}

ICRS &	International Celestial Reference System \\
FK5 &	Mean Place, New (IAU 1984) System \\
FK4 &	Mean Place, Old (Bessell-Newcomb) System \\
FK4-NO-E	mean & Place, Old System but without e-terms \\
GAPPT &	Geocentric Apparent Place, IAU 1984 System \\

\end{tabular}
\end{center}
\caption{RA-DEC systems recognised by \Rwcs, and by extension \propane.}
\label{tab:ra_dec}
\end{table}%


\bsp	
\label{lastpage}
\end{document}